\documentclass[preprint2]{aastex}

\newfont{\boldit}{cmbxti10}
\DeclareMathAlphabet{\mitbf}{OML}{cmm}{b}{it}

\usepackage{subfig}
\usepackage{amsmath}

\slugcomment{Accepted for publication on ApJ,  March 22th 2010}

\shorttitle{CARMA observations of RY Tau and DG Tau.}
\shortauthors{Isella A. et al.}

\begin{document}


\title{Investigating planet formation in circumstellar disks:\\ 
CARMA observations of RY Tau and DG Tau.}
 

\author{Andrea Isella, John M. Carpenter, Anneila I. Sargent}
\affil{Division of Physics, Mathematics and Astronomy, California Institute of Technology, MC 249-17, Pasadena, CA 91125.}
\email{isella@astro.caltech.edu}


\begin{abstract}
We present CARMA observations of the thermal dust emission from the circumstellar 
disks around the young stars RY~Tau and DG~Tau at 
wavelengths of 1.3~mm and 2.8~mm. The angular resolution of the maps is as 
high as 0.15\arcsec, or 20 AU at the distance of the Taurus cloud, which is  a factor of 2 
higher than has been achieved to date at these
wavelengths. The unprecedented detail of the resulting disk images
enables us to address three important questions related to 
the formation of planets. (1) What is the radial distribution of 
the circumstellar dust?  (2) Does the dust emission show any 
indication of gaps that might signify the presence of 
(proto-)planets? (3) Do the dust properties depend on the orbital radius?

We find that modeling the disk surface density in terms of either a classical 
power law or the similarity solution for viscous disk evolution, reproduces 
the observations well. Both models constrain the surface density between 15 
and 50 AU to within 30\% for a given dust opacity. Outside this range, the densities inferred 
from the two models differ by almost an order of magnitude. The 1.3 mm image from RY~Tau
shows two peaks separated by 0.2\arcsec\ with a decline in the dust emission toward the stellar position,
which is significant at about 2-4$\sigma$.
For both RY~Tau and DG~Tau, the dust emission at radii larger than 15 AU displays 
no significant deviation from an unperturbed viscous  disk 
model. In particular, no radial gaps in the dust distribution are detected. 
Under reasonable assumptions, we exclude the presence 
of planets more massive than 5 Jupiter masses orbiting either star at distances 
between about 10 and 60 AU,  unless such a planet is so young that there has been 
insufficient time to open a gap in the disk surface density. The radial variation 
of the dust opacity slope, $\beta$,  was investigated by comparing 
the 1.3 mm and 2.8 mm observations.  We find mean values 
of $\beta$ of 0.5 and 0.7 for DG~Tau and RY~Tau respectively.
Variations in $\beta$ are smaller than $\Delta\beta=0.7$ between 
20 and 70 AU. These results confirm that the circumstellar dust throughout  
these disks differs significantly from dust in the interstellar 
medium.

\end{abstract}


\keywords{}

\section{Introduction}

Resolved images of circumstellar disks around young stars provide the 
most direct tool for investigating the formation of planets. At 
millimeter wavelengths, the thermal dust emission is generally optically 
thin and measures the radial distribution of circumstellar dust 
\citep{bs90}. However, since circumstellar disks in nearby star forming 
regions typically have radii between 100 and 500 AU, sub-arcsecond angular 
resolution is required to spatially resolve the dust emission, even in nearby 
star-forming clouds. Millimeter-wave interferometers are essential for such studies.

Since sub-arcsecond observations at millimeter wavelengths require both high 
sensitivity and high dynamical range, only a small number of bright disks have 
been observed at resolutions of 0.4\arcsec-1\arcsec\ to date \citep{b08,gu99,is07,pi05,pi06,
pi07,sim00,t03,wi00}. The Combined Array for Research in 
Millimeter-wave Astronomy (CARMA) and the new extended configuration 
of the Sub-Millimeter Array are rapidly enabling more extensive  
high resolution surveys of circumstellar disks, particularly in the 
Taurus and Ophiuchus star forming regions \citep[][hereafter Paper I]{an09,hu09,is09}. 

The highest angular resolution achieved so far by millimeter-wave interferometers 
is 0.3\arcsec-0.4\arcsec, corresponding to spatial scales of 40-50~AU at 
the distance of Taurus and Ophiuchus. In most cases, the dust density 
appears to increase smoothly inward down to the orbital radius 
resolved by the observations, typically $\sim$25~AU. However, central cavities 
in the dust distribution are revealed in a number of disks \citep{an09,hu09}. 
It remains a matter of debate whether these cavities are caused by dynamical
interactions, inside out disk dispersal mechanisms, dust 
opacity variations, or viscous evolution \citep[e.g.,][Paper I]{al06,c05,cm07,dd05}. 

Nevertheless, these observations still lack the angular 
resolution required to resolve the innermost part of the disk where the 
density of the circumstellar material is highest and 
the formation of planets is more probable. Here we describe 
CARMA observations of the thermal dust emission towards the young 
stars DG~Tau and RY~Tau at an angular resolution of 0.15\arcsec\ 
at 1.3~mm and 0.3\arcsec\ at 2.8~mm. At the distance of 
Taurus (140~pc), 0.15\arcsec\ corresponds to spatial scales 
of 20~AU, such that emission on orbital scales comparable to Saturn can 
be resolved. This is more than a factor of two improvement over previous 
observations of circumstellar disks at these wavelengths. 

DG~Tau and RY~Tau are classical T Tauri stars of spectral type M0 
and K1 respectively \citep{mu98,kh95}. Stellar ages inferred from 
stellar evolutionary models are less than 
1~Myr (see Paper I for more details and references). The relative 
youth of both systems is confirmed by the presence of 
large amounts of gas and dust extending to 0.1~pc and by associated 
stellar jets and outflows \citep[see, e.g,][]{mr04,sb08}. From 
near-infrared to millimeter wavelengths, both objects exhibit
strong  emission in excess of that from the stellar photospheres. 
This is attributed to rotating disks with radii of few hundred 
AU that first absorb and then re-emit radiation from the central stars.
\citep{ks95,t02}. Our earlier CARMA observations of 1.3~mm 
thermal dust emission from these disks, at a resolution of 0.7\arcsec, 
suggested disk masses between 5 and 150\% of the stellar 
mass for both sources (see Paper I). These high 
disk masses and the youth of RY~Tau and 
DG~Tau make these prime targets to investigate the earliest 
stages of planet formation. Our new observations of RY~Tau and DG~Tau
have a factor of 5 better angular resolution and a factor of 3 better sensitivity
than the previous data. 

This paper investigates three main questions related to the formation of 
planets in young circumstellar disks. (1) What is the surface density distribution
in the observed disks down to an orbital radius of 10 AU? (2) Are there any 
signatures of planet formation contained in the dust distribution?  Finally, (3), do the dust properties 
vary with orbital radius? A qualitative answer to the first two questions is 
proposed in Section~\ref{sec:morp} where we present the observations and discuss 
the morphology of the dust emission. A quantitative analysis is described 
in Section~\ref{sec:mod}, where we compare the observations with theoretical 
models of disk emission. Implications of these results for 
disk structure, for the possible presence of planets, and for the radial
variation of the dust opacity are considered 
in Section~\ref{sec:res}. The conclusions are presented 
in Section~\ref{sec:conc}.

\section{Observations and data reduction}
\label{sec:obs}

We observed thermal dust emission from the RY~Tau and DG~Tau 
circumstellar disks using CARMA in the A, B, and C configurations. 
The date of observation, 
array configurations used, baseline range, sizes and 
orientations of the synthesized beams, integrated fluxes, 
seeing and noise levels are summarized in Table~\ref{tab:obs}. 
The C-configuration observations 
were presented in Paper I. 

\begin{deluxetable}{llllllll}
\tablecaption{Summary of CARMA continuum observations \label{tab:obs}}
\tablewidth{\textwidth}
\tablehead{
\colhead{Object} & \colhead{Date} & \colhead{Array} & \colhead{Baseline} &\colhead{Beam}  
& \colhead{Flux}  & \colhead{Seeing} & \colhead{Noise} \\ 
\colhead{}       & \colhead{(UTC)}     & \colhead{Configuration}        & \colhead{range (m)}       & \colhead{FWHM(\arcsec), PA(\arcdeg)} & \colhead{(mJy)} & \colhead{(\arcsec)} & \colhead{(mJy/beam)}
}
\startdata
\multicolumn{6}{l}{Observations at 1.3 mm} \\
\hline
DG~Tau  \rule{0pt}{3.0ex} &             & A+B+C &         & 0.17$\times$0.15, 103 &  367$\pm$14 & 0.03  & 0.96  \\
 ...    \rule{0pt}{1.0ex} & {\small 2007 Oct 08} & {\small C} & {\small 21-279}   & {\small 1.04$\times$0.81, 112} &  & \\ 
 ...     & {\small 2007 Dec 14} & {\small B} & {\small 81-937}   & {\small 0.33$\times$0.29, 116} &  & \\
 ...     & {\small 2009 Jan 31} & {\small A} & {\small 130-1884} & {\small 0.15$\times$0.13, 102} &  & \\
RY~Tau  \rule{0pt}{3.0ex} &             & A+B+C &         & 0.17$\times$0.14, 81 &  227$\pm$7 & 0  & 0.90 \\ 
...     \rule{0pt}{1.0ex} & {\small 2007 Oct 22} & {\small C} & {\small 16-280}    & {\small 1.24$\times$0.78, 102} &  & \\
...      & {\small 2008 Dec 30} & {\small B} & {\small 82-935}    & {\small 0.40$\times$0.31, 109} &  & \\
...      & {\small 2009 Jan 19} & {\small A} & {\small 139-1884}  & {\small 0.15$\times$0.13,  82} &  & \\
\hline
\multicolumn{6}{l}{\normalsize Observation at 2.7 mm} \rule{0pt}{3.0ex} \\
\hline
DG~Tau \rule{0pt}{3.0ex} &             & A+B &          & 0.45$\times$0.38, 131  & 58$\pm$6 & 0.07 & 0.45 \\
...    \rule{0pt}{1.0ex} & {\small 2008 Jan 14} & {\small B} & {\small 80-798}   & {\small 1.12$\times$0.60, 125} &  &\\
...     & {\small 2009 Feb 05} & {\small A} & {\small 136-1678} & {\small 0.34$\times$0.32, 163} &   & \\

RY~Tau  \rule{0pt}{3.0ex}&             & A+B &          & 0.36$\times$0.30, 82 & 36$\pm$3 & 0.07 & 0.28 \\       
...     \rule{0pt}{1.0ex}& {\small 2008 Feb 01} & {\small B} & {\small 82-945}    & {\small 0.77$\times$0.59, 95}  &  &\\
...     & {\small 2009 Feb 10} & {\small A} & {\small 123-1884}  & {\small 0.35$\times$0.29, 82}  &  & \\
\enddata

\end{deluxetable}

The observations were obtained at LO frequencies of 228.1 
GHz ($\lambda=1.3$~mm) and 106.2 GHz ($\lambda=2.8$~mm). 
The CARMA correlator at the time of the observations contained three bands, 
each of which was configured to 468 MHz bandwidth to provide maximum continuum 
sensitivity. The band pass shape was calibrated 
by observing 3C273; flux calibration was set by observing 
Uranus and 3C84. The radio galaxy 3C111 was observed every 9 minutes
to correct for atmospheric and instrumental effects. Variations of 
the atmospheric conditions on time scales shorter than 9 minutes 
are not corrected, in effect resulting in seeing. 
We quantified the atmospheric seeing by measuring the size of the phase 
calibrator image; if the seeing is negligible the phase 
calibrator appears as a point source. Otherwise, the seeing produces 
a Gaussian smoothing that can be quantified through the full width half 
maximum (FWHM) of the resulting image. We find that at 1.3~mm 
the effect of seeing is negligible for RY~Tau but produces a FWHM 
of 0.03\arcsec\ for DG~Tau. Atmospheric conditions were slightly worse 
during the 2.8~mm observations, resulting in seeing of 0.07$\arcsec$ for 
both objects. These seeing estimates do not account for variations in the
 atmospheric conditions on angular scales of 10\arcdeg, 
 corresponding to the separation between the source and 
the calibrator.  Values for the atmospheric seeing are summarized in 
Table~\ref{tab:obs} and are adopted in the model fitting described in 
Section~\ref{sec:mod}.

\begin{figure*}[!t]
\centering
\includegraphics[angle=0,width=0.8\textwidth]{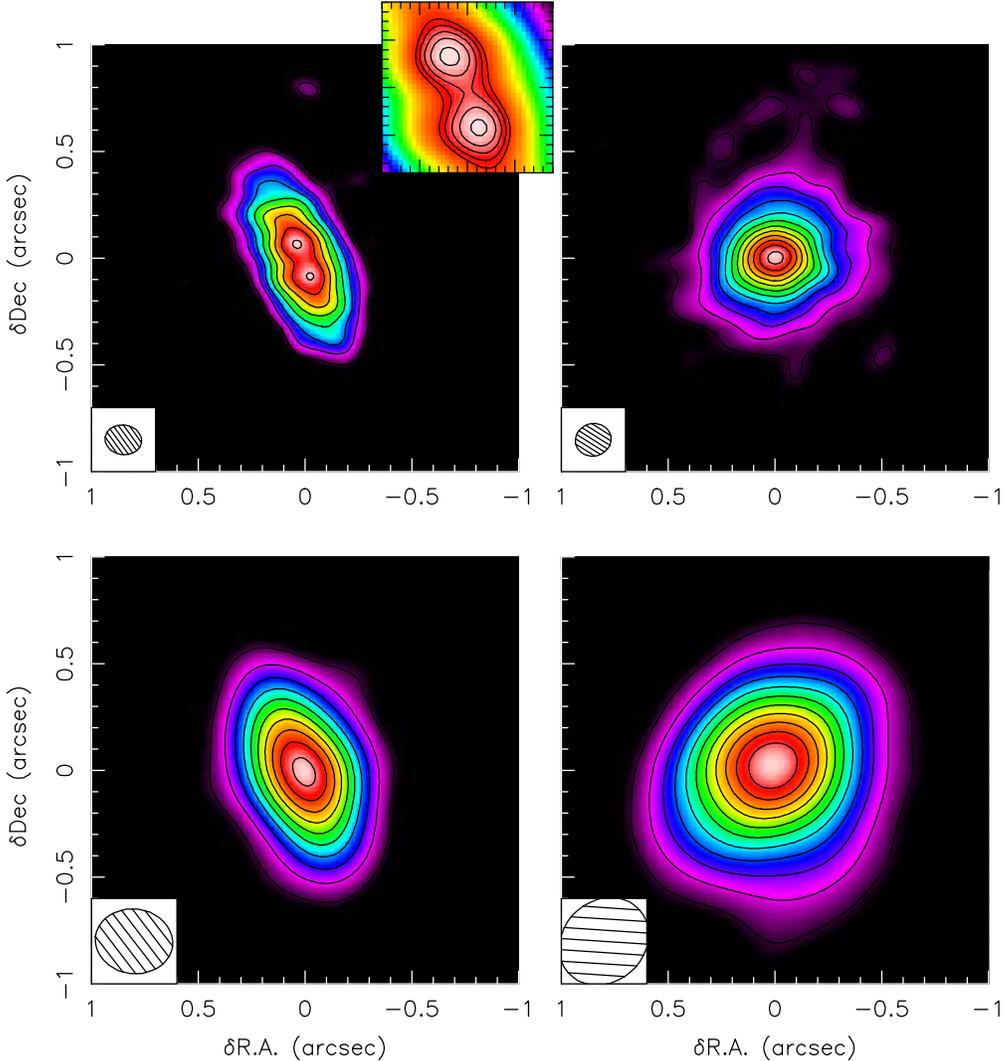}
\caption{\label{fig:cont} Maps of the dust thermal emission observed at a
wavelength of 1.3~mm (upper panels) and 2.8 mm (lower panels) towards 
RY~Tau (left panels) and DG~Tau (right panels). The color scale shows the 
surface brightness starting from the $3\sigma$ level, with contours 
plotted every 4$\sigma$. The 1$\sigma$ noise level and the size of the 
synthesized beam are given in Table~\ref{tab:obs}.  The inset in the 
upper left panel shows the central 0.4\arcsec$\times$0.4\arcsec\ region of 
the RY~Tau disk where contours start at 28$\sigma$ with increments of 1$\sigma$. 
The surface brightness is characterized by two peaks separated 
by $\sim$0.2\arcsec. 
}
\end{figure*}

\begin{figure*}[!t]
\centering
\includegraphics[angle=0,width=0.9\textwidth]{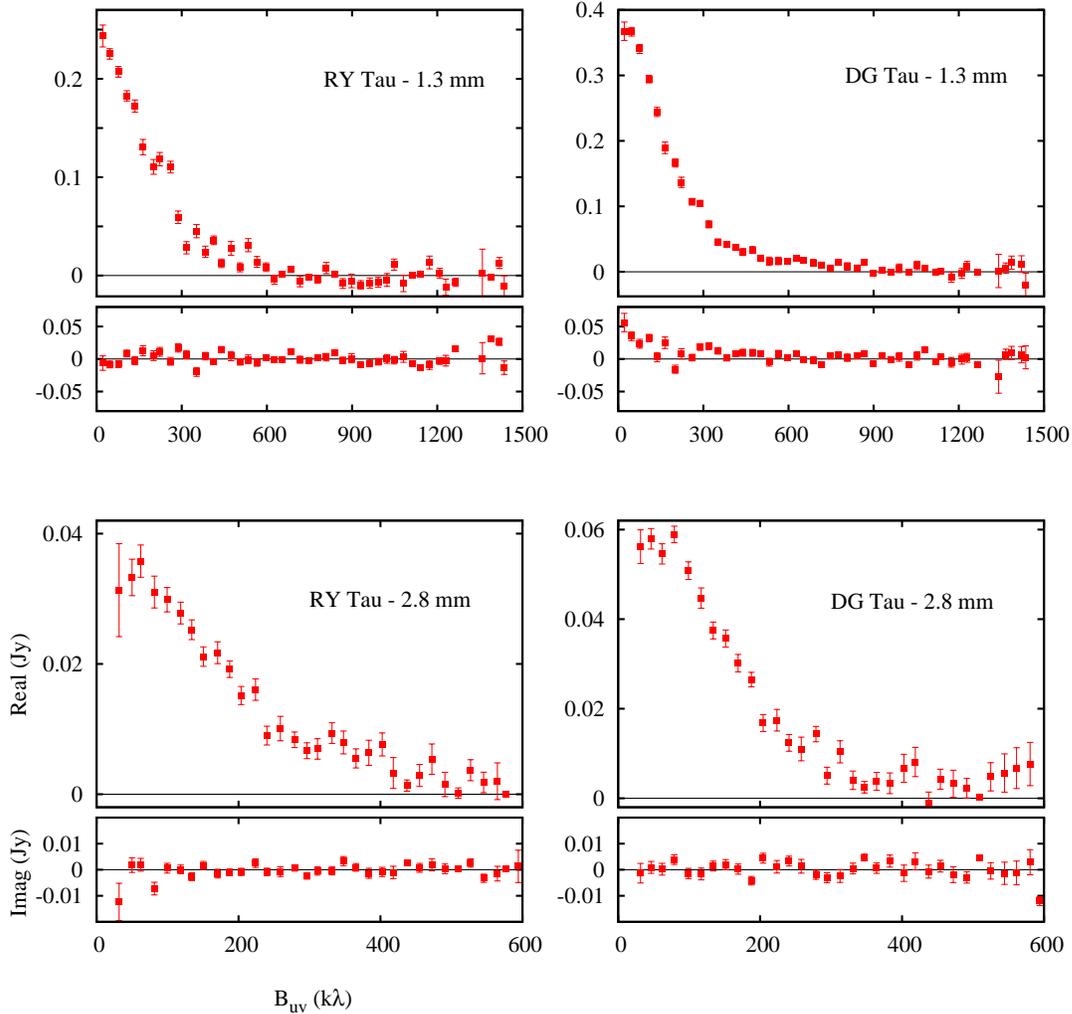}
\caption{\label{fig:uvamp} Correlated flux measured at 1.3~mm and 2.8~mm 
toward RY~Tau (left panels) and DG~Tau (right panels) as function of the baseline 
length $B_{uv}$ expressed in k$\lambda$. }
\end{figure*}

The raw data were reduced using the MIRIAD software package. The maps of 
the continuum emission shown in Figure~\ref{fig:cont} were derived using 
GILDAS software. Corresponding   
complex visibilities are shown in Figure~\ref{fig:uvamp}. At 1.3 mm, natural 
weighting of the A, B and C configuration observations produced a FWHM synthesized beam 
size of $\sim$ 0.15\arcsec. The noise levels are 0.96~mJy/beam and 0.90~mJy/beam respectively
for DG~Tau and RY~Tau. Dust emission at 2.8~mm was observed in the A and B 
configurations at angular 
resolution of $\sim$ 0.35\arcsec\ and noise 
levels of 0.45 mJy/beam and 0.28 mJy/beam for DG~Tau and RY~Tau respectively. 

\section{Morphology of the dust emission}
\label{sec:morp}
In Figure~\ref{fig:cont}, the dust emission in both disks is 
clearly resolved and characterized by a smooth and centrally 
symmetric radial profile. DG~Tau intensity contours are almost 
circular suggesting a disk inclination smaller than 30\arcdeg. 
For RY~Tau, the intensity contours are elongated in the 
North-East direction suggesting a disk position angle of 
about 24\arcdeg\ measured East from North and a disk inclination 
of at least 65\arcdeg. For both sources the disk orientations agree 
with those found in Paper I.

\subsection{RY Tau disk morphology}
 
\begin{figure}[!t]
\centering
\includegraphics[angle=0,width=\columnwidth]{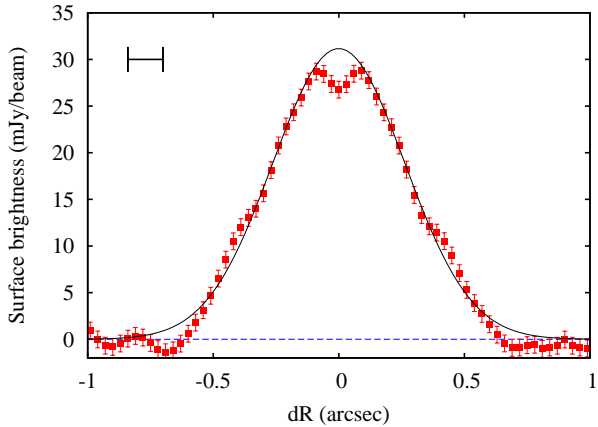}
\caption{\label{fig:radialprofile} Radial profile of the 1.3 mm 
surface brightness in RY Tau disk measured along the major axis of the disk at a
position angle of 24\arcdeg\ East from North.  The intensity error bars (red points)
correspond to the noise level of 0.9 mJy/beam. The solid curve shows a gaussian fit, while 
the dashed line indicates the zero intensity level. The bracket in the upper left 
indicates the angular resolution. }
\end{figure}

The 1.3~mm dust continuum emission from the RY~Tau disk shows 
two spatially resolved peaks separated by about 0.2\arcsec (28~AU), and 
oriented along the apparent major axis of the disk. Details of the central 
0.4\arcsec$\times$0.4\arcsec\ region are displayed in the inset in the upper 
left panel of Figure~\ref{fig:cont}, and the radial profile of the surface 
density along the disk major axis is shown in Figure~\ref{fig:radialprofile}.
The intensity at both peaks is 29 mJy/beam, which is  2 mJy/beam (i.e. 2.2$\sigma$) 
higher than the intensity at the center of the disk. We also estimated the expected 
central surface brightness by fitting a gaussian to the surface brightness distribution 
at angular distances larger than 0.15\arcsec. The fitted gaussian is shown as the 
solid curve in Figure~\ref{fig:radialprofile}.  A gaussian function was chosen since it 
provides a reasonable parametric representation of the dust emission. Interpolating 
this gaussian fit to the center of the disk suggests an expected central surface 
brightness of 31 mJy/beam, which is 4$\sigma$ higher than the measured value.
The significance level of the two intensity peaks, the fact that they appear 
in the map before cleaning, their orientation along the disk major axis, and the symmetry with respect 
to the central star, suggest that they are real and, therefore, that the dust emission decreases 
inside an orbital radius of about 14~AU. This is analogous to the situation in "transitional'' disks, 
where the inner gaps observed  in the dust emission are attributed to dusty depleted inner 
regions \citep[see, e.g.,][]{hu09,b08,b09}. 

At a first sight, this interpretation is incompatible 
with RY~Tau's large near- and mid-infrared excesses, which suggest the 
presence of  warm dust within 10 AU of the star \citep{ro07}. If, however, the 
inner disk is only partially depleted and dust emission remains optically thick 
in the infrared, the observed double intensity peak and the spectral energy distribution 
 can be reconciled. A number of physical mechanisms could reduce the dust density in the inner 
region of circumstellar disks. For example,  planets less massive than Jupiter
may carve partially depleted gaps in the surface density distribution by tidal 
interaction with the surrounding material \citep{br99}. This possibility is discussed in 
more detail in Section~\ref{sec:surf}.  In Paper I, we  also proposed that a surface 
density profile that gradually decreases towards the star may originate naturally 
from the viscous evolution of a disk if viscosity decreases with radius. Finally, it
is also possible that the decrease in dust emission may originate from a lowering of
 opacity due to the growth of dust grains to centimeter sizes \citep{dd05}. 
 Unambiguously disentangling these models requires even higher 
angular resolution observations than are yet available.
 
We must note that radial velocities studies \citep{hb98} and Hipparcos observations 
of the variability of the photocenter \citep{be99} suggest that 
RY~Tau is a binary. The Hipparcos data implies a minimum projected 
separation of 23.6 mas and a position angle of 304\arcdeg$\pm$34\arcdeg, 
almost perpendicular to the position angle of the disk inferred 
from our observations. Assuming that the binary and the disk have 
the same inclination, the spatial separation between 
the binary components is 6-9 AU, and could explain the double peak in the dust 
continuum emission. Indeed, the presence of a stellar mass 
companion orbiting at a radius of 6-9~AU would push the inner radius of 
the circumstellar disk to a distance of 9-13 AU by tidal interactions 
\citep{wo07}. However, the binary nature of RY~Tau has been rendered 
questionable by near-infrared interferometric observations 
that suggest an inner disk radius at 0.1~AU from the central 
star and exclude the presence of a stellar mass companion between 
0.35 AU and 4 AU down to a stellar flux ratio of 0.05 \citep{ak05,pp09}.
A stellar companion was also undetected in recent spectroscopic 
and aperture masking observations 
(Duy Cuong Nguyen and Adam Kraus private communication). As 
discussed above,  the spectral energy distribution is  also inconsistent with the 
existence of a large inner gap completely depleted of gas and dust as could be 
expected for a stellar companion \citep{ro07}. 
These results suggest that RY~Tau is indeed a single star, and 
the variability observed by Hipparcos and the radial velocity variations may be attributed 
to brightness changes in the circumstellar environment \citep[see the discussion in][and references therein]{sc08}.

A notable characteristic of our images of the dust emission is the 
high degree of central symmetry and, with the exception of the
innermost region, the almost complete absence of 
features in the surface brightness distribution. If the emission is optically 
thin (we will examine this assumption is Section~\ref{sec:surf}), this 
translates to a smooth radial profile for the dust. The degree of symmetry of the 
emission can be quantified by analyzing the imaginary part of the correlated flux, 
plotted in Figure~\ref{fig:uvamp} as a function of the angular frequency $B_{uv}$. 
Point symmetric emission will have a zero imaginary part at all spatial frequencies. 
For RY~Tau, the deviations from zero 
are comparable to the noise in the observations (see the left panels of Figure~\ref{fig:uvamp})

\subsection{DG Tau disk morphology}

The surface brightness distribution for the DG~Tau disk has a
central peak coincident with the stellar position and smoothly decreases 
outward to reach the noise level at an angular distance of about 
0.5\arcsec. At both 1.3 and
2.8 mm the emission appears fairly symmetric, and indeed the imaginary part 
of the correlated flux is zero for $B_{uv}>300$~k$\lambda$ (see 
upper right panel of Figure~\ref{fig:uvamp}). The imaginary part increases 
up to 50 mJy at shorter spatial frequencies, suggesting that the 
emission may be asymmetric on angular scales larger than  
$1/B_{uv}\sim 0.7$\arcsec.  Comparing the real and imaginary parts of 
the correlated flux at the shortest spatial frequencies, we find 
that the asymmetric part of the emission contributes about 14\% to the total 
flux. 


As already noted, the high angular resolution observations 
of RY~Tau and DG~Tau allow us to examine disk properties 
that bear on planet formation. In particular the
radial density profile of the circumstellar material 
is critical to understanding where planet 
formation may occur, or where it has occurred.
With observations at more than one wavelength, we may also 
consider radial variations of the grain properties.
As we describe below, measurements of the dust radial profile and the variation 
of the dust properties with radius are best undertaken by comparing the 
observations with theoretical disk models in the Fourier domain. There, 
the effects of different angular resolutions, sensitivity and atmospheric 
seeing can be more easily taken into account. Moreover, comparison with 
theoretical models is the only way to quantify the contribution from 
optically thick emission, leading to an improved 
estimate of the dust mass. 


\section{Disk and dust  models}
\label{sec:mod}

\begin{deluxetable}{lllll}
\tablecaption{Properties of the adopted dust models \label{tab:dust}}
\tablewidth{\columnwidth}
\tablehead{
\colhead{Object} & \colhead{$a_{max}$} & \colhead{$q$} & \colhead{$k_{1.3~mm}$}  & \colhead{$k_{2.8~mm}$}\\
                 & \colhead{(cm)}     &               & \colhead{(cm$^2$/g)}   & \colhead{(cm$^2$/g)}   
}
\startdata
\hline
\multicolumn{4}{l}{High dust opacity model, $H$}\\ 
DG~Tau & 0.075 & 3 & 0.082 & 0.056 \\ 
RY~Tau & 0.035 & 3 & 0.131 & 0.075 \\ 
\hline
\multicolumn{4}{l}{Low dust opacity model, $L$}\\ 
DG~Tau & 5.0 & 3.5 & 0.012 & 0.0078 \\ 
RY~Tau & 5.0 & 3.7 & 0.026 & 0.015   
\enddata
\end{deluxetable}

To investigate the dust radial distribution around DG~Tau and RY~Tau, we
consider two different models for the disk surface density. The first model consists of 
the classical power law parameterization 
\begin{equation}
\label{eq:pow}
\Sigma(R) = \Sigma_{40} \left( \frac{R}{40\textrm{AU}} \right)^{-p}\, \textrm{for} \,\,  R_{in}<R<R_{out},
\end{equation} 
where $\Sigma_{40}$ is the disk surface density at a radius of 40~AU. 
$R_{in}$  and $R_{out}$ are the inner and 
outer disk radii respectively. The second model is the similarity solution for 
the evolution of a viscous Keplerian disk \citep{lb74,har98}. 
As discussed in Paper I, this has the form 
\begin{multline}
\label{eq:sim}
\Sigma(R,t) = \Sigma_t \left( \frac{R_t}{R} \right)^{\gamma} \\ \times  \exp{ \left\{ - \frac{1}{2(2-\gamma)} 
\left[ \left( \frac{R}{R_t} \right)^{(2-\gamma)} -1 \right] \right\} }
\end{multline}
where $\Sigma_t$ is the surface density at radius $R_t$, sometimes called the 
{\it transition radius}. For $R \lesssim R_t$, the surface density has
a power law profile characterized by the slope $\gamma$, while at
larger radii the surface density falls exponentially. 
 
These two different parameterizations are used to calculate the dust 
emission by solving the structure of an hydrostatic disk heated by the stellar 
radiation \citep{ddn01}. The temperature on the disk mid plane is self-consistently 
calculated by adopting the {\it two layer} approximation of  \citet{cg97}.  The disk 
temperature, which depends mainly on the disk optical depth at optical and infrared wavelengths, 
is roughly proportional to $R^{-1/2}$ for both surface density profiles 
(see Paper I and reference therein for a detailed discussion on the disk 
temperature radial profile). 

Fundamental to any disk model is the prescription adopted 
for the dust opacity. Although the dust contributes only 
about 1\% to the total disk mass, it dominates the disk opacity 
in the wavelength range where most of the stellar and disk radiation is emitted. 
We assume that the dust size distribution follows a 
power law $n(a) \propto a^{-q}$ where $a$ is the radius of a dust grain. 
The assumptions on the slope $q$, on the minimum and maximum grain 
sizes, on the dust chemical composition and on the grain structure define the frequency 
dependence of the dust opacity $k_{\nu}$, and, ultimately, the disk emission. 
The dust opacity is calculated for compact spheres composed of astronomical 
silicates and organic carbonates \citep{wd01,zu96}. We assume a
mass ratio of 1 between silicates and organics, which leads to grain density of 
2.5 g/cm$^3$. The dust opacity averaged over the grain size distribution 
is calculated by fixing the minimum grain size to 0.005 $\mu$m. The maximum 
grain size $a_{max}$ and the slope $q$ are set to reproduce 
the observed slope of the spectral energy distribution as discussed below.

At millimeter wavelengths the dust opacity can be approximated by
a power law $k_{\nu}=k_0(\nu/\nu_0)^\beta$ \citep{bs91}.
If the dust emission is optically thin and the Rayleigh-Jeans 
approximation is satisfied, the slope $\beta$ of the dust opacity is 
related to the spectral index $\alpha$ of the observed disk 
emission $F_{\nu}$ ($F_\nu \propto \nu^\alpha$) by the
relation $\alpha=2+\beta$. This relation is only approximate if the 
dust emission is optically thick at some radii. In Paper I, 
we derived values for $\beta$ of 0.5 and 0.7 for DG~Tau and RY~Tau 
respectively from an analysis of the SED, taking into account the 
optically thick contribution to the total dust emission. For the assumed 
dust composition and structure, these values of $\beta$ can be reproduced 
with different choices of the maximum grain size $a_{max}$ and the grain 
size slope $q$ (see Appendix~\ref{app:A}). 
To investigate how the assumptions on the grain size distribution affect the 
model fitting, we adopt  two different dust models that correspond to the 
extreme cases of low ($L$) and high ($H$) opacity.  The corresponding dust 
opacities at both 1.3~mm and 2.8~mm are given in Table~\ref{tab:dust}.  

Finally, we assume that the dust opacity is constant 
throughout the disk. This is indeed one of the main assumption we want to test by 
modeling the observed dust emission at 1.3~mm and 2.8~mm and will be discussed 
in detail in Section~\ref{sec:betavar}.\\
\vspace{0.1 cm}

\section{Results and Discussion}
\label{sec:res}

Models and observations are compared in Fourier space to avoid the 
non linear effects introduced by the cleaning process. 
The best fit models are found by $\chi^2$ minimization with five free 
parameters: the disk inclination $i$,  the disk position angle PA, $R_{out}$, $\Sigma_{40}$, and $p$ 
for the  power law surface density (Equation~\ref{eq:pow}), and $i$, PA, 
$R_t$, $\Sigma_t$ and $\gamma$ for the similarity 
solution (Equation~\ref{eq:sim}). The disk inner radius $R_{in}$ is 
fixed at 0.1~AU.  For both surface density models we find best 
fit solutions for both the high ($H$) and low ($L$) dust opacity 
models. The 1.3~mm and 2.8~mm data are fitted independently. 

To minimize $\chi^2$ and evaluate the constraints on the model parameters, 
we use a Bayesian approach that adopts uniform prior probability distributions. 
In practice we sample the $\chi^2$ probability distribution by varying the free parameters using the Markov Chain Monte Carlo 
method described in Paper I.

\begin{deluxetable}{llllllll}
\tablecolumns{8}
\tablecaption{Best fit parameters assuming $H$ dust model. \label{tab:res_clubs}}
\tablewidth{\textwidth}
\tablehead{
\multicolumn{8}{c}{Similarity solution} \\
\cline{1-8} 
\colhead{Object} & \colhead{$\lambda$ (mm)} & \colhead{$i$ (\arcdeg)}   & \colhead{PA (\arcdeg)}     & \colhead{$R_t$ (AU)} & \colhead{$\gamma$} & \colhead{$\Sigma_t$ (g/cm$^2$)} &  \colhead{$\chi^2_r$}  
}

\startdata
DG~Tau  \rule{0pt}{3.0ex} & 1.3 & 24$\pm$9  & 119$\pm$23 & 23.4$\pm$1.8 & 0.33$\pm$0.15  & 10.9$\pm$1.5 & 1.0608 \\
...                       & 2.8 & 31$\pm$12 & 144$\pm$19 & 27.7$\pm$3.0 & 0.10$\pm$0.24  & 7.5$\pm$1.3  & 1.0629 \\
RYTau	\rule{0pt}{3.0ex} & 1.3 & 66$\pm$2  &  24$\pm$3  & 26.7$\pm$1.2 & -0.54$\pm$0.18 & 2.6$\pm$0.2  & 1.0896 \\
...                       & 2.8 & 71$\pm$6  &  20$\pm$4  & 26.5$\pm$2.7 & -0.08$\pm$0.54 & 2.6$\pm$0.5  & 1.1894 \\
\cutinhead{Power law}
Object & $\lambda$ (mm) & $i$ (\arcdeg)  & PA (\arcdeg)   & $R_{out}$ (AU)  & p & $\Sigma_{40}$ (g/cm$^2$) & $\chi^2_r$ \\
\hline
DG~Tau  \rule{0pt}{3.0ex} & 1.3 & 27$\pm$8  & 120$\pm$26 & 72.6$\pm$6.3  & 1.00$\pm$0.15  & 5.6$\pm$1.5  & 1.0611 \\
...                       & 2.8 & 32$\pm$11 & 144$\pm$18 & 82.2$\pm$10.5 & 0.74$\pm$0.24  & 4.5$\pm$1.6  & 1.0629 \\
RYTau	\rule{0pt}{3.0ex} & 1.3 & 66$\pm$2  & 24$\pm$3   & 70.6$\pm$3.9  & 0.12$\pm$0.15  & 1.9$\pm$0.6  & 1.0897 \\
...                       & 2.8 & 71$\pm$6  & 20$\pm$4   & 76.9$\pm$12.0 & 0.64$\pm$0.45  & 1.6$\pm$1.0  & 1.1894 \\
\enddata
\tablecomments{The uncertainties correspond to a likelihood of 99.7\% (i.e. 3$\sigma$)
for the normal distributions shown in Figure~\ref{fig:RYTau_POW}-\ref{fig:DGTau_SIM}.}
\end{deluxetable}

Once a best fit solution is found, we confirm that
this indeed corresponds to an absolute minimum of $\chi^2$, as opposed to a 
local minimum, by running multiple Monte Carlo simulations with random initializations 
and verifying that they all converge to the same solution. Each parameter
is allowed to vary in a large range: 0-80\arcdeg\ for the inclination, $\pm$90\arcdeg\ 
for the position angle, $10-1000$~AU for $R_t$ and $R_{out}$, $\pm4$ for $p$ and $\gamma$,
and $0.1-1000$~g/cm$^2$ for $\Sigma_{40}$ and $\Sigma_t$.   

\begin{figure*}[!t]
\centering
\includegraphics[angle=0,width=0.8\textwidth]{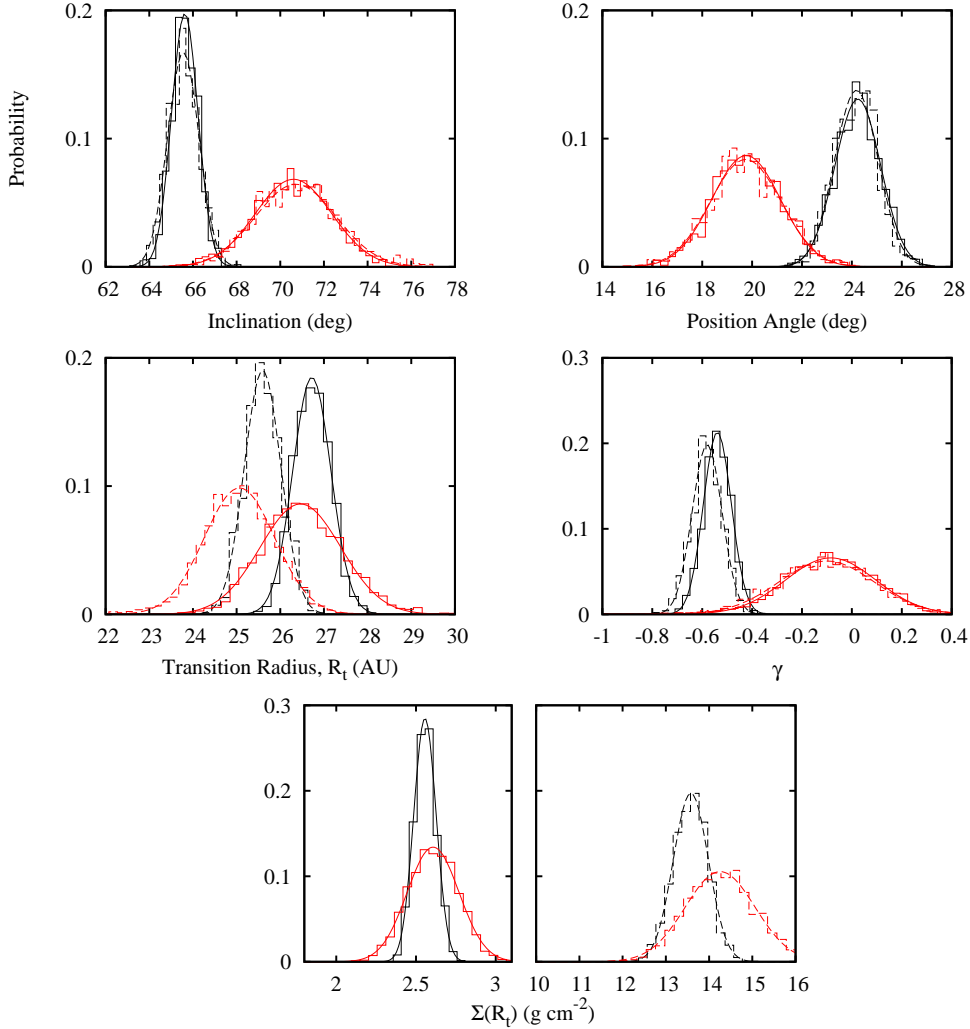}
\caption{\label{fig:RYTau_SIM} Probability distribution of the disk parameters obtained by fitting  the RY~Tau
observations at 1.3~mm and 2.8~mm with the similarity solution for 
the surface density distribution. The results obtained by fitting the 1.3~mm data only are shown by the black curves, 
while the red curves indicate the results obtained by fitting the 2.8~mm only. Solid curves show the probability distribution 
obtained assuming the dust opacity model $H$, and dashed curves correspond to the dust opacity model $L$ (see Table~\ref{tab:dust}).
}
\end{figure*}

\begin{figure*}[!t]
\centering
\includegraphics[angle=0,width=0.8\textwidth]{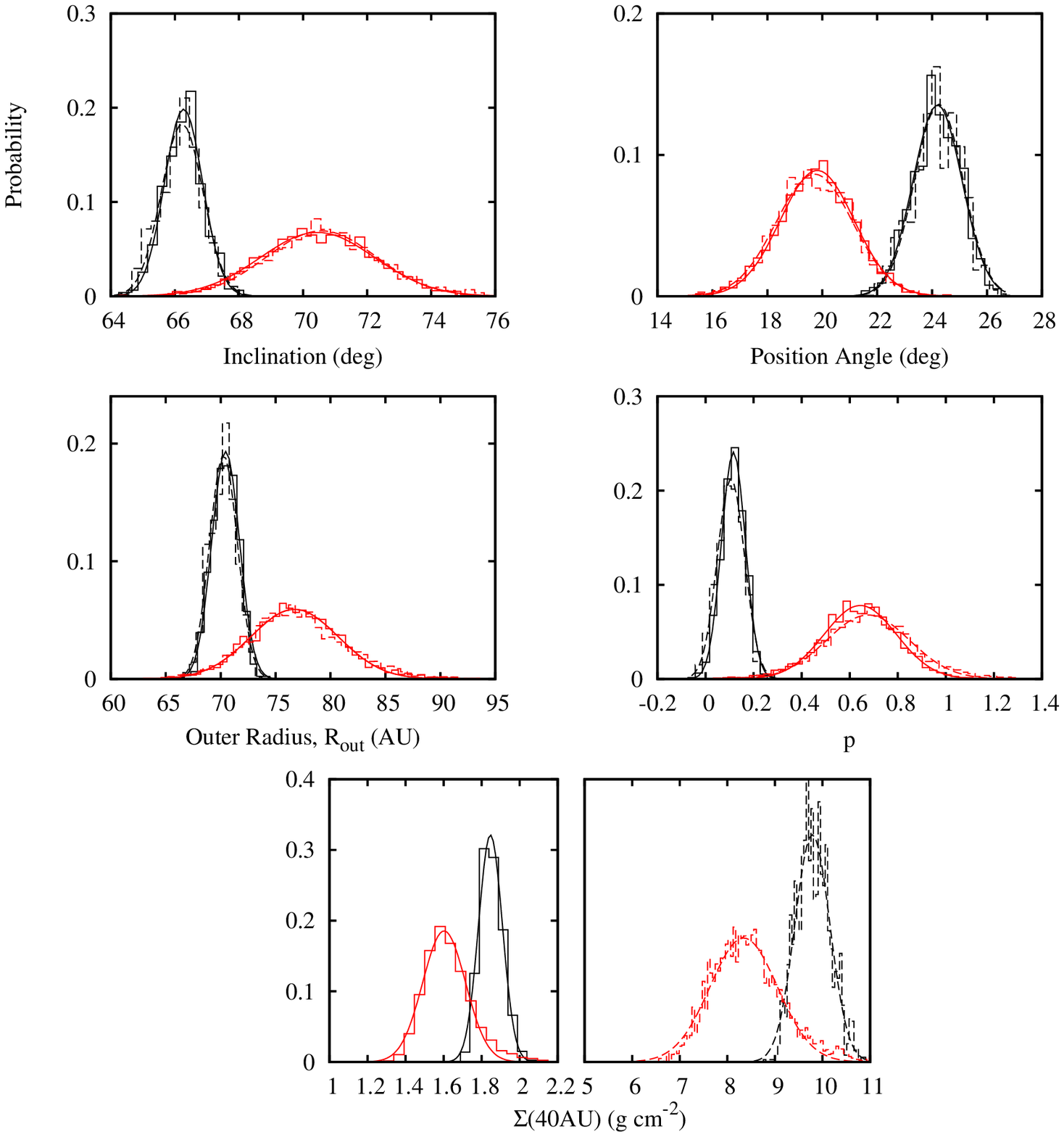}
\caption{\label{fig:RYTau_POW} Probability distribution of the disk parameters obtained by fitting the 
RY~Tau observations at 1.3~mm and 2.8~mm with a power law surface density 
distribution. Colors, solid curves, and dashed curves are the same as in Figure~\ref{fig:RYTau_SIM}.
}
\end{figure*}

\begin{figure*}[!t]
\centering
\includegraphics[angle=0,width=0.8\textwidth]{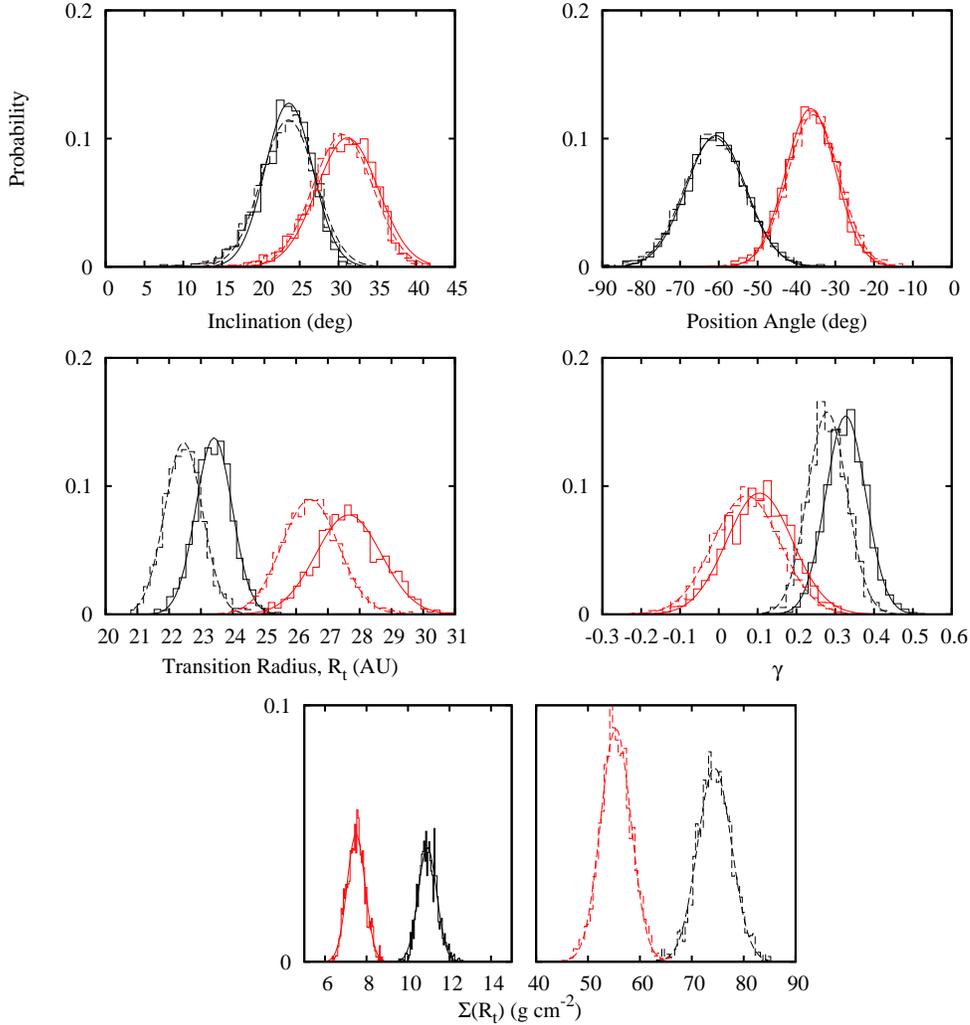}
\caption{\label{fig:DGTau_SIM} Probability distribution of the disk parameters obtained by fitting the DG~Tau
observations at 1.3~mm and 2.8~mm with the similarity solution for 
the surface density distribution. Colors, solid curves, and dashed curves are the same as in Figure~\ref{fig:RYTau_SIM}.
}
\end{figure*}

\begin{figure*}[!t]
\centering
\includegraphics[angle=0,width=0.8\textwidth]{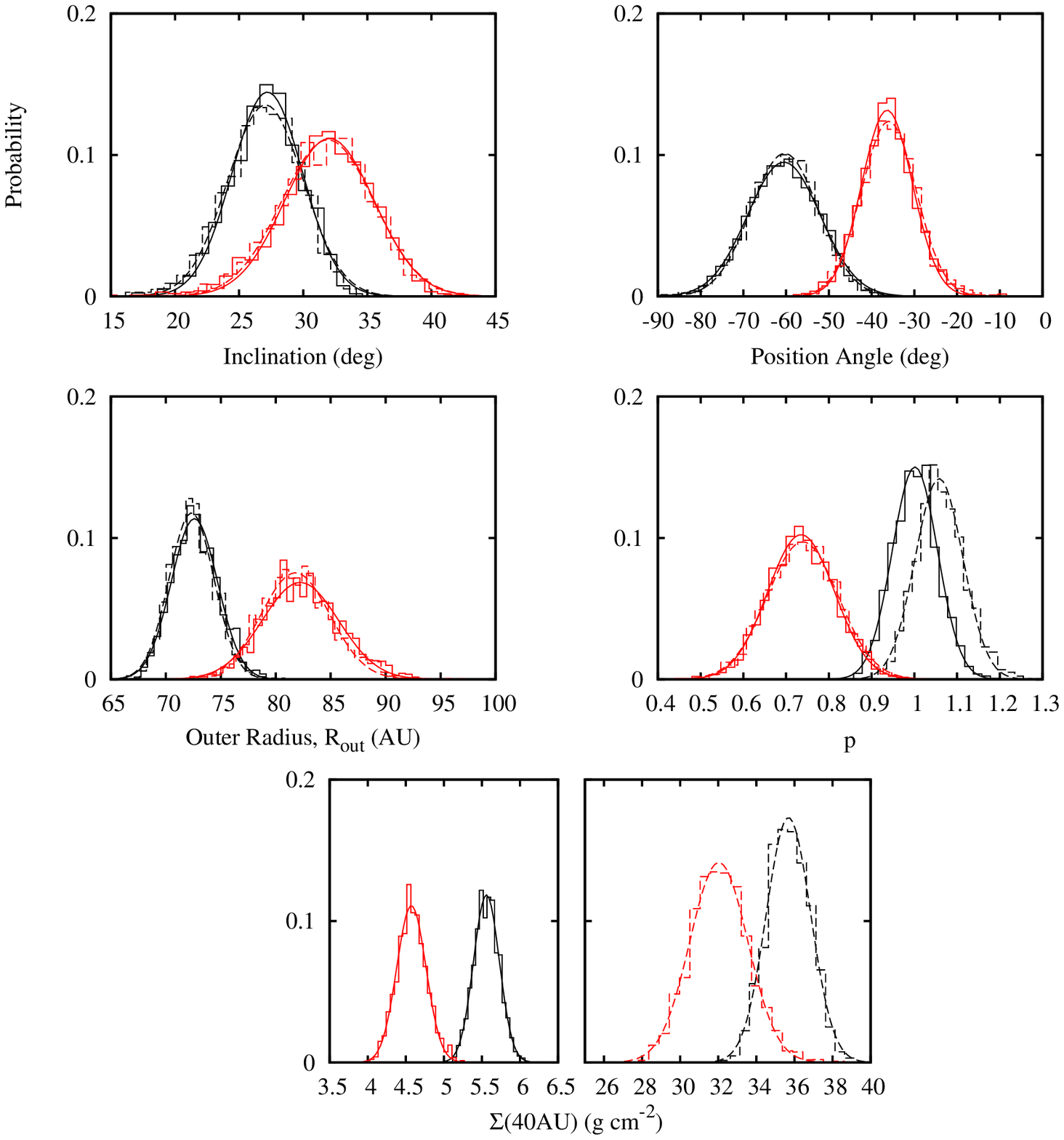}
\caption{\label{fig:DGTau_POW} Probability distribution of the disk parameters obtained by fitting the DG~Tau
observations at 1.3~mm and 2.8~mm with a power law surface density distribution. 
Colors, solid curves, and dashed curves are the same as in Figure~\ref{fig:RYTau_SIM}.
}
\end{figure*}

The best fit disk models found for high and low dust opacities are listed 
in Table~\ref{tab:res_clubs} and Table~\ref{tab:res_spades} respectively. 
Each table lists the parameters for the similarity 
solution disk model in the upper part, and for the power law disk model in the 
lower part. The probability distributions for each free parameter are shown 
in Figure~\ref{fig:RYTau_SIM}  and Figure~\ref{fig:RYTau_POW} for RY Tau in the case of 
the similarity solution and power law respectively. The same quantities for DG~Tau 
are shown in  Figure~\ref{fig:DGTau_SIM} and Figure~\ref{fig:DGTau_POW}.
In these figures, the black and red histograms indicate the probability distributions
derived by fitting the 1.3~mm and 2.8~mm observations, respectively; 
solid and dashed curves represent the $H$ and $L$ dust opacity models. 
For each parameter we derive the uncertainty range that corresponds 
to a likelihood of 99.7\% (3$\sigma$) 
by fitting a normal distribution to the 
probabilities.

\begin{figure*}[!t]
\centering
\includegraphics[angle=0,width=1.0\textwidth]{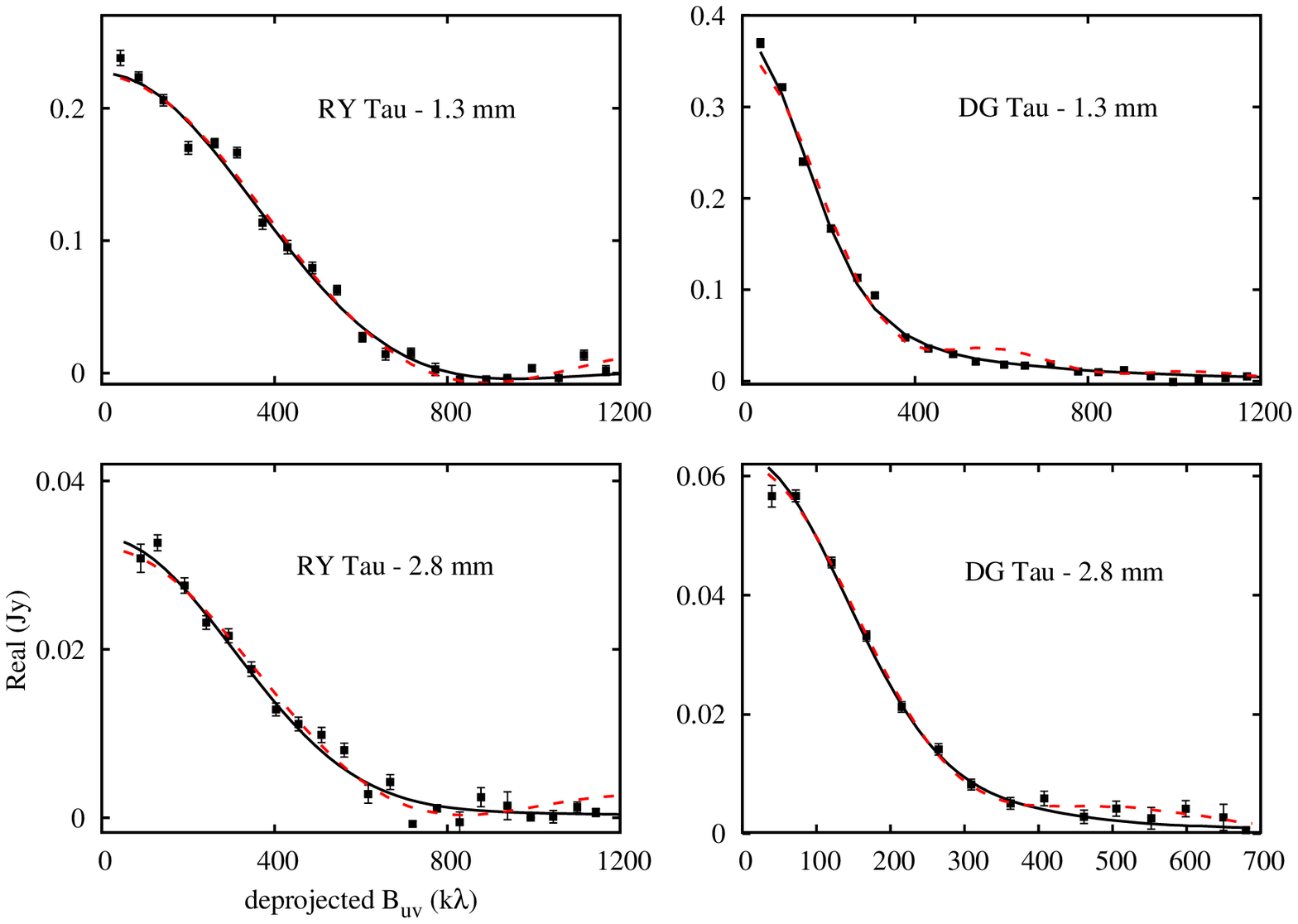}
\caption{\label{fig:uvamp_deproj_real} The real part of the correlated flux 
as function of the deprojected baseline. The solid curves show the best-fit models where 
the surface density is described by the similarity solution for viscous disk evolution.  
The dashed curves show the results for a model with a power law surface density.
The models were computed using the high dust opacity, but identical results were obtained for the low 
dust opacity case.}
\end{figure*}

Finally, Figure~\ref{fig:uvamp_deproj_real} shows comparisons 
between the observed real part of the correlated flux (filled 
squares with error bars), the best fit models for the similarity 
solution (solid curve), and a power law surface density (dashed curve).

\begin{deluxetable}{llllllll}
\tablecolumns{8}
\tablecaption{Best fit parameters assuming $L$ dust model. \label{tab:res_spades}}
\tablewidth{\textwidth}
\tablehead{
\multicolumn{8}{c}{Similarity solution} \\
\cline{1-8} 
\colhead{Object} & \colhead{$\lambda$ (mm)} & \colhead{$i$ (\arcdeg)}   & \colhead{PA (\arcdeg)}    & \colhead{$R_t$ (AU)} & \colhead{$\gamma$} & \colhead{$\Sigma_t$ (g/cm$^2$)} &  \colhead{$\chi^2_r$}   
}
\startdata
DG~Tau  \rule{0pt}{3.0ex} & 1.3 & 24$\pm$11 & 119$\pm$24 & 22.5$\pm$1.8 &  0.28$\pm$0.15 & 74.4$\pm$9.9 & 1.0608\\
...                       & 2.8 & 31$\pm$12 & 144$\pm$20 & 26.4$\pm$2.7 &  0.07$\pm$0.27 & 55.4$\pm$8.9 & 1.0629\\
RYTau	\rule{0pt}{3.0ex} & 1.3 & 66$\pm$2  &  24$\pm$3 & 25.6$\pm$1.2 & -0.58$\pm$0.18 & 13.6$\pm$1.2 & 1.0896\\
...                       & 2.8 & 71$\pm$6  &  20$\pm$4 & 25.1$\pm$2.4 & -0.10$\pm$0.57 & 14.3$\pm$2.3 & 1.1893\\
\cutinhead{Power law}
Object & $\lambda$ (mm) & $i$ (\arcdeg)   & PA (\arcdeg)   & $R_{out}$ (AU)  & p & $\Sigma_{40}$ (g/cm$^2$) & $\chi^2_r$  \\     
\hline
DG~Tau  \rule{0pt}{3.0ex} & 1.3 & 27$\pm$9  & 120$\pm$24  & 72.3$\pm$4.0 & 1.06$\pm$0.18  & 35.7$\pm$3.6 & 1.0611\\
...                       & 2.8 & 32$\pm$11 & 144$\pm$19  & 81.8$\pm$9.3 & 0.74$\pm$0.24  & 32.1$\pm$4.5 & 1.0629\\
RYTau	\rule{0pt}{3.0ex} & 1.3 & 66$\pm$2  &  24$\pm$3  & 70.5$\pm$3.9 & 0.11$\pm$0.18   & 9.7$\pm$1.2  & 1.0897\\
...                       & 2.8 & 71$\pm$5  &  20$\pm$4  & 76.7$\pm$12.6 & 0.68$\pm$0.51   & 8.3$\pm$2.3  & 1.1893\\
\enddata
\tablecomments{The uncertainties correspond to a likelihood of 99.7\% (i.e. 3$\sigma$)
for the normal distributions shown in Figure~\ref{fig:RYTau_POW}-\ref{fig:DGTau_SIM}.}
\end{deluxetable}

\vspace{0.2 cm}

\subsection{Dependence on the dust opacity and implications on the 
disk masses}
\label{sec:res_dust}
The best fit solutions for the $H$ and $L$ dust opacity models 
are shown in Figure~\ref{fig:RYTau_POW}-\ref{fig:DGTau_SIM} with solid 
and dashed curves respectively. In all cases, $H$ and $L$ models lead to 
very similar values for the disk position angle, the disk inclination 
and the radial profiles of the surface density defined by $p$ and $R_{out}$ 
in the case of the power law models, and $\gamma$ and $R_t$ for the 
similarity solution models. As discussed in Paper I, 
these parameters are essentially independent of the dust opacity. 
This is mainly because the disk mid-plane temperature $T_i(R)$ varies by only a 
few percent between the different dust models, as long as the disk is 
optically thick to the stellar radiation. Since $\Sigma(R) \propto T_i(R)^{-1}$, 
the radial profile of the surface density varies by only small fraction when 
different dust models are assumed.

By contrast, the surface density normalization ($\Sigma_t$ and $\Sigma_{40}$) 
varies with the dust opacity so that the product $\Sigma \times k_{\nu}$ 
remains almost constant if the emission is optically thin. Consequently, 
a lower dust opacity requires a higher dust mass in order to emit the same 
amount of radiation at millimeter wavelengths. The ratio 
$\Sigma_{L}/\Sigma_{H}$ is then approximately equal to the ratio between 
the dust opacities listed in Table~\ref{tab:dust}.

From the analysis of the surface density of the best fit model we find that 
the RY~Tau emission is always optically thin at both 1.3 and 2.8~mm. However, 
DG~Tau emission is optically thick within 20~AU at 1.3 mm for both the 
similarity solution and the power law models. The 1.3~mm flux emitted within 
this region is about 25\% of the total flux. At 2.8~mm the emission
is always optically thin in the case of the similarity solution while it 
is optically thick within 6~AU in the power law case. In this case the 
optically thick contribution is 5\% of the total flux.

Different dust opacities lead to different values for the total mass 
of dust in the disks. For DG~Tau we obtain total 
dust masses of about 33 and 233 Earth masses ($M_{\earth}$) in the 
case of the high and low opacity dust models respectively. 
Disk masses of $\sim$10 and 50~$M_{\earth}$ are found for 
RY~Tau. 
Massive 
disks can also be obtained by extending the grain size distribution 
larger than 5~cm. For example, in Paper I we derived total dust 
masses of about 1331 and 216 $M_{\earth}$ for 
DG~Tau and RY~Tau respectively by assuming a maximum grain size of 10~cm 
and a slightly different grain composition. 
Additional uncertainties in the disk mass come from the  dust chemical composition. 
As discussed in Appendix~\ref{app:A}, the presence 
of ice or vacuum in the grains leads to smaller dust opacities at millimeter 
wavelengths and consequently produces higher disk masses. 
We therefore estimate that the circumstellar disks 
around DG~Tau and RY~Tau contain a minimum mass of dust of 
30 and 10 $M_{\earth}$ respectively, while the upper limit is not 
constrained due to the uncertainties on the grain size distribution. 
Assuming the standard dust/gas ratio of 0.01, these values correspond to minimum 
disk masses of  0.009 and 0.003 $M_{\sun}$ for the DG~Tau and RY~Tau respectively. 

\begin{figure}[h]
\centering
\includegraphics[angle=0,width=0.9\columnwidth]{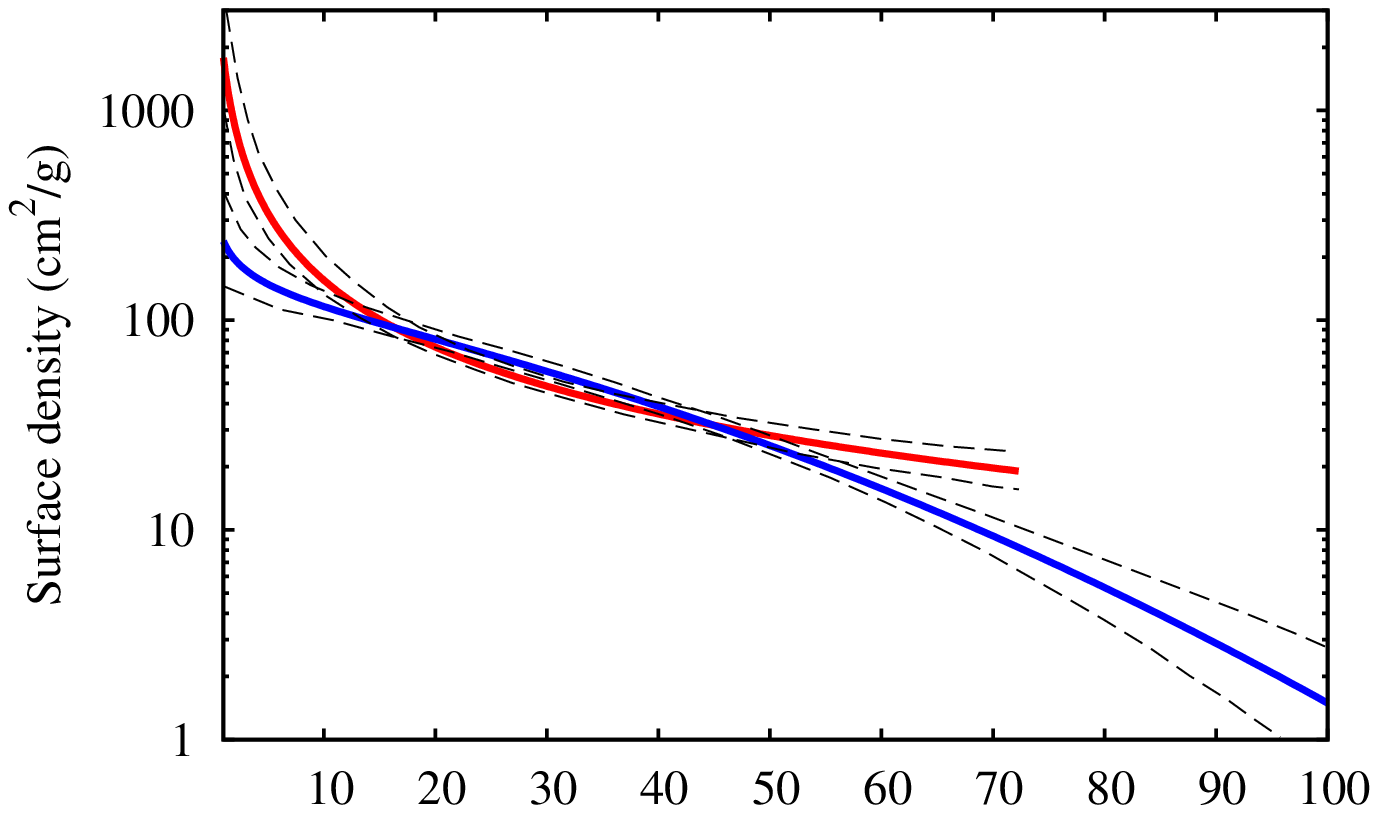}
\includegraphics[angle=0,width=0.9\columnwidth]{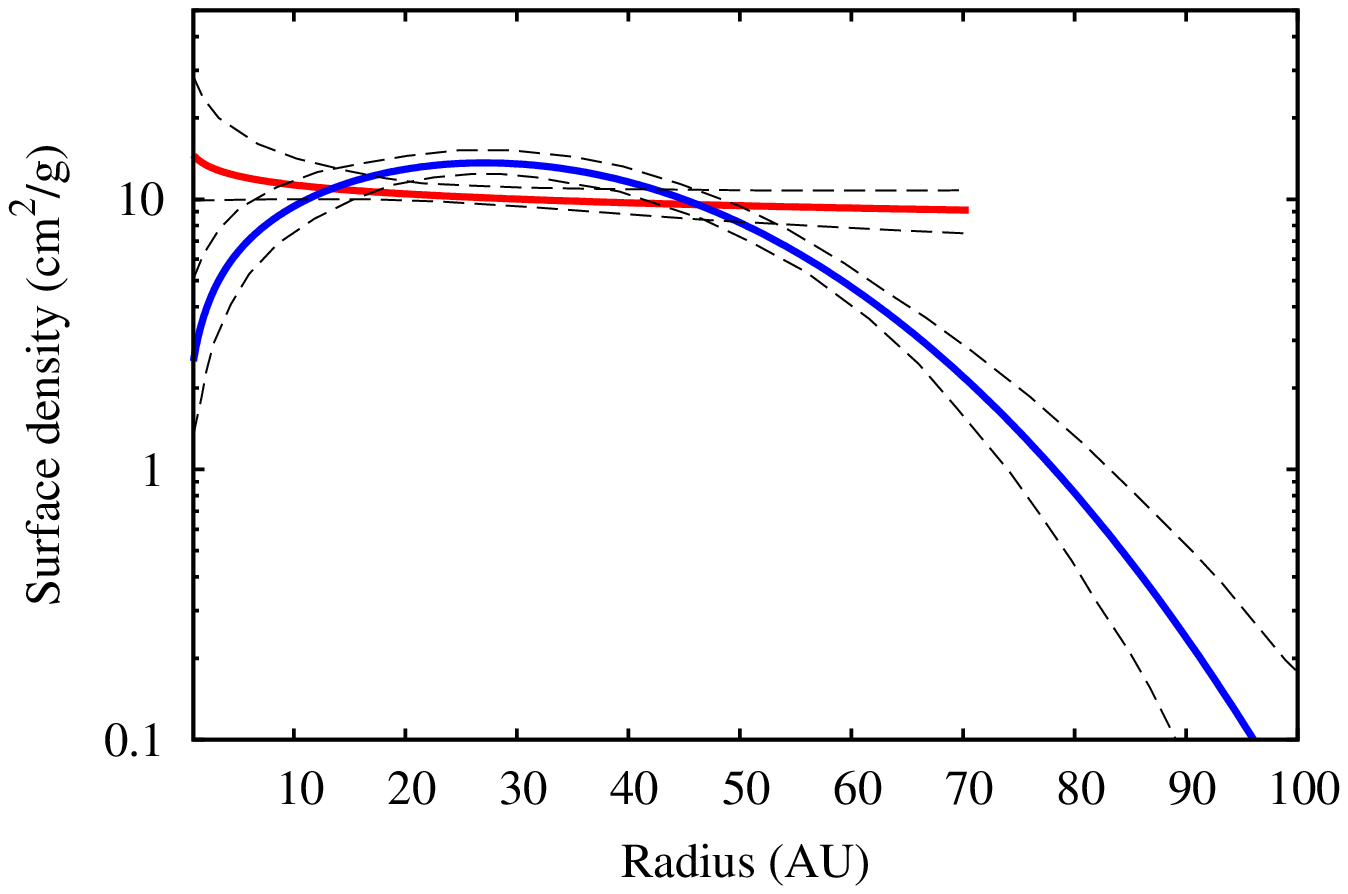}
\caption{\label{fig:sigma} Surface density for DG Tau (top panel) and RY~Tau 
(lower) disk. The red and blue curves show the best fit solutions for the power law  and the 
similarity solution models respectively in the low opacity case $L$. 
The dashed curves show the 3$\sigma$ uncertainty range for the surface density.
The surface densities for the high opacity case will have similar shapes, but 
will be about a factor of 6 lower. 
}
\end{figure}

\subsection{Constraints on the surface density: similarity solution vs power law}
\label{sec:sim}
Figure~\ref{fig:uvamp_deproj_real} shows the comparison between models 
and observations in terms of the real part of the correlated flux as a 
function of the baseline length. To correct for the disk inclination we
deprojected the baseline assuming the inclinations and position angles listed 
in Table~\ref{tab:res_clubs}. In this figure, the results for $H$ and $L$ dust 
models lead to indistinguishable curves. Similarity solution and power law 
models are represented with solid and dashed curves respectively, and the 
observations are shown by black dots with error bars. It is  clear that both 
the similarity solution and power law disk models provide satisfactory fits 
to the observations. The similarity solution model provides smaller 
values of $\chi^2$ (see Table \ref{tab:res_clubs} and \ref{tab:res_spades}) 
and, in the case of DG Tau, a better fit to the 
observations between 400 and 800 k$\lambda$. In this range of spatial frequencies, 
the power law solution is characterized
by a wiggle due to the sharp truncation of the dust emission at 72~AU.
On the other hand, the exponential tapering of the similarity solution 
leads to a smooth visibility profile that matches extremely well 
the observations. The same behavior is present in the lower panel 
which compares the model and the observations at 2.8~mm. However, 
in this case the observations at $B_{uv} > 400 $k$\lambda$ are too
to distinguish between the two models. 
Although not conclusive, this result make the similarity solution model 
a more appealing explanation for the dust emission in circumstellar disks, 
confirming the conclusions of \citet{hu08}. 

Figure~\ref{fig:sigma} shows the surface density derived from the 1.3~mm
observations for both the power law and the similarity 
solution model in the case of high dust opacity. The two models 
lead to similar values of $\Sigma(R)$ in the region where most of 
the 1.3 mm flux is emitted, namely between $\sim$15 and 50 AU. 
In this region, the surface density in RY~Tau disk is almost 
constant with the radius, while it decreases roughly as $1/R$ in 
the case of DG~Tau. 
Inside 15 AU and outside 50 AU, the observations lack both 
the angular resolution and the sensitivity required to directly 
constrain the surface density. As a consequence, the values of 
$\Sigma(R)$ strongly depend on the assumed model and can
differ by one order of magnitude at the disk inner radius. 
 
\subsection{Surface density and implication on the existence of planets}
\label{sec:surf}

In this section we discuss the implications of the inferred 
surface density on the presence of planets. The 
analysis is limited to surface density profiles 
obtained by fitting the observations at 1.3~mm, which 
have the highest angular resolution.

\subsubsection{DG~Tau}

For the similarity solution model, the surface density 
has a radial profile characterized by $\gamma \sim 
0.31\pm0.18$ and $R_t\sim 23\pm2$~AU. The transition radius 
$R_t$ agrees well with our earlier observations (21$\pm$3) 
but $\gamma$ is significantly larger than the value 
of -0.5$\pm$0.6 from Paper I. The 
discrepancy is probably due to the fact that the earlier observations
were taken in poorer weather conditions and the model fitting did not account 
for the atmospheric seeing. Figure 10 shows the residuals 
after subtracting the best fit model to the new observations. Note that the 
power law model gives very similar residuals. The residuals are as high 
as 3-6$\sigma$  and are found at angular scales larger
than 0.7\arcsec\ where the emission is slightly asymmetric 
(see Section~\ref{sec:morp}). In this outermost disk region, the surface 
density may deviate significantly from the symmetric radial profile 
assumed in the model. We calculate that variations of 
$\pm 10-30$ g/cm$^2$ with respect to the best fit surface density profile over a 
spatial region comparable with the beam size may produce the observed residuals. 
Larger variations of the surface density on smaller angular scales are also 
possible.
\begin{figure}[t]
\centering
\includegraphics[angle=0,width=\columnwidth]{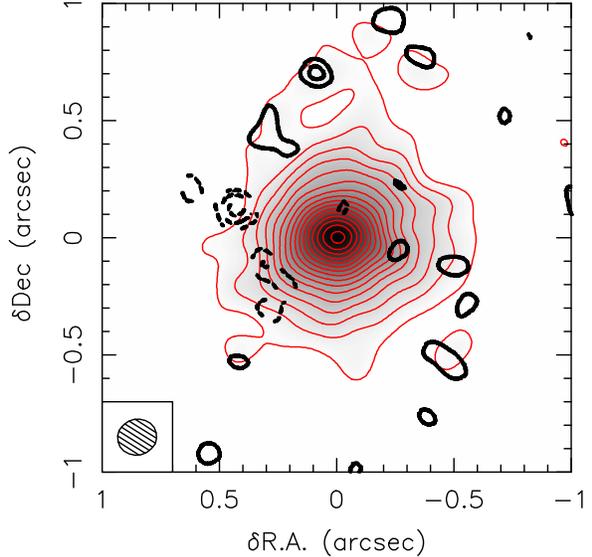}
\caption{\label{fig:res_DGTau} The black curves show the residuals for 
the 1.3 mm DG~Tau observations after subtracting the best-fit model for the similarity 
solution. Contours start at 3$\sigma$ and are 
spaced by 1$\sigma$. The thin red curves and the color scale 
show the observed dust emission, with contours spaced by 3$\sigma$.
}
\end{figure}

The residuals do not show global deviations from the smooth surface 
density profile, apparently excluding the possibility of gaps in the 
dust distribution that might be produced by a planet. Of course, low mass 
planets may not produce any discernible gap and may still exist in the DG~Tau disk. 
The formation of a gap is possible only if the efficiency 
in removing the material close to the planet orbital radius via tidal torques 
is larger than the mass accretion rate due to the disk viscosity 
\citep[see, e.g,][]{lp93}. If we assume the $\alpha$ prescription for the disk 
viscosity \citep{ss73}, and call $h$ the pressure scale height of the 
disk, a planet orbiting at radius $R_p$ can open a radial 
gap in the disk surface density only if $M_p/M_\star > 32 \alpha (h/R_p)^2$ 
\citep{lp93,br99}. Moreover, for a disk in hydrostatic 
equilibrium with the gravitational field of the central star, the pressure scale 
$h$ is proportional to  $T_i(R_p)^{1/2} R_p^{3/2} M_{\star}^{-1/2}$ \citep{cg97}. Since the 
temperature is $T_i(R_p) \propto R_p^{-1/2}$ (see Paper I for more details),
the formation of a gap requires
\begin{equation}
M_p > 26.3 R_p^{1/2} \alpha,
\end{equation}
where $M_p$ is expressed in Jupiter masses, $R_p$ is AU, and the numerical constant
is calculated for a disk temperature of 194 K at 1~AU as determined from 
our disk model. Typical values of $\alpha$ are in the range $10^{-2}-10^{-3}$, 
and imply that a planet can open a gap at 1~AU only if its mass is larger
than about 0.1~$M_J$. To open a gap at 30~AU, the mass must be larger than 
about 0.5~$M_J$. 

To investigate the effects that a planet more massive than 
0.1 $M_J$ might have on the observations of the dust continuum emission, we simulated 
the presence of a planet in the DG~Tau disk by opening a gap in the surface density 
distribution corresponding to the best fit models discussed above. 
For simplicity, we assumed that the planet describes a circular orbit and that 
the gap can be represented by a circular ring. To be compatible with numerical simulations of 
planet-disk interaction, the half-width of the ring $\Delta$ is assumed to be equal 
to twice the Hill radius $R_H = R_p \sqrt[3]{M_P/(3M_{\star})}$ 
\citep[e.g.][]{br99,wo07}. In the region between $R_p \pm \Delta$
the surface density is depleted by a fraction $f$ that depends on the mass of 
the planet and on the disk viscosity. For $\alpha = 10^{-3}$, we can assume 
$f=0$ for planet masses M$_{p} > 1 M_J$, $f=0.1$ for $M_p=0.5$~$M_J$, $f=0.17$ 
for $M_p=0.3~M_J$ and  $f=0.6$ for $M_p=0.1~M_J$ \citep{wo07}. Therefore, 
only planets more massive than 1~$M_J$ will produce completely cleaned gaps. 

We simulated gaps corresponding to planets in the mass 
range 0.3-5~$M_J$ and with orbital radii between 1 and 90 AU. For each model we 
calculated the residuals as the difference between the observations of DG~Tau 
at 1.3~mm and the model image. If the gap is too small compared to our angular 
resolution, or too faint compared with our sensitivity, the residuals will be similar 
to the case without gaps shown in Figure~\ref{fig:res_DGTau}. In this case we 
say that the gap is not detected. On the other hand, a large and deep gap
 will produce a bright ring in the residual. To quantify how reliable the 
detection of a gap is, we define a signal-to-noise ratio of the gap, {\it gap SNR},
in the following way. First, we deproject the residual for the inclination and 
position angle of DG~Tau disk. Then we take the radial average of the residuals
at the distance corresponding to the orbital radius of the planet adopting a radial bin width equal 
to the FWHM of the synthesized beam (i.e., 0.17\arcsec).  We define  
the {\it gap SNR} as the mean residual in the radial bin divided by the uncertainty in the mean.
In this way, detected gaps correspond to {\it gap SNR}$> 3$. 

\begin{figure}[!ht]
\centering
\includegraphics[angle=0, width=\columnwidth]{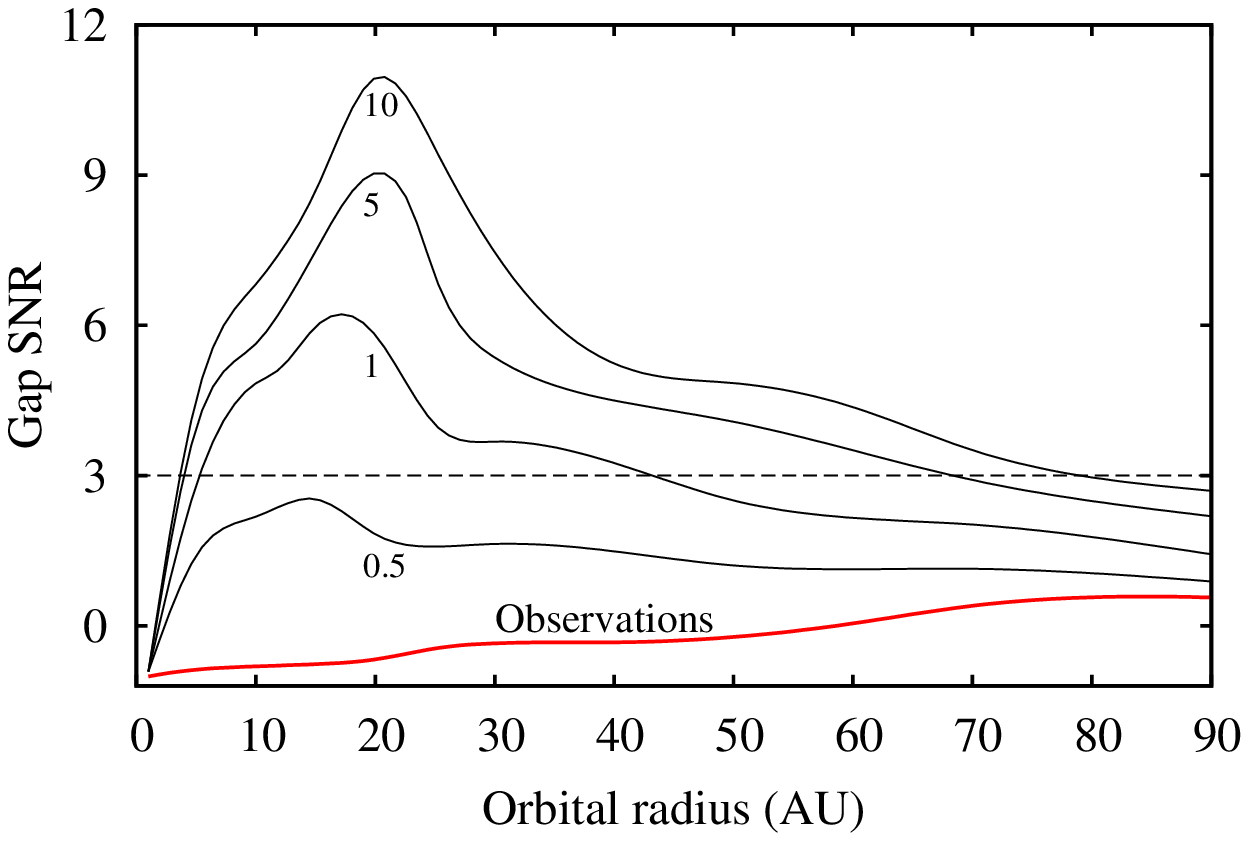}
\includegraphics[angle=0, width=\columnwidth]{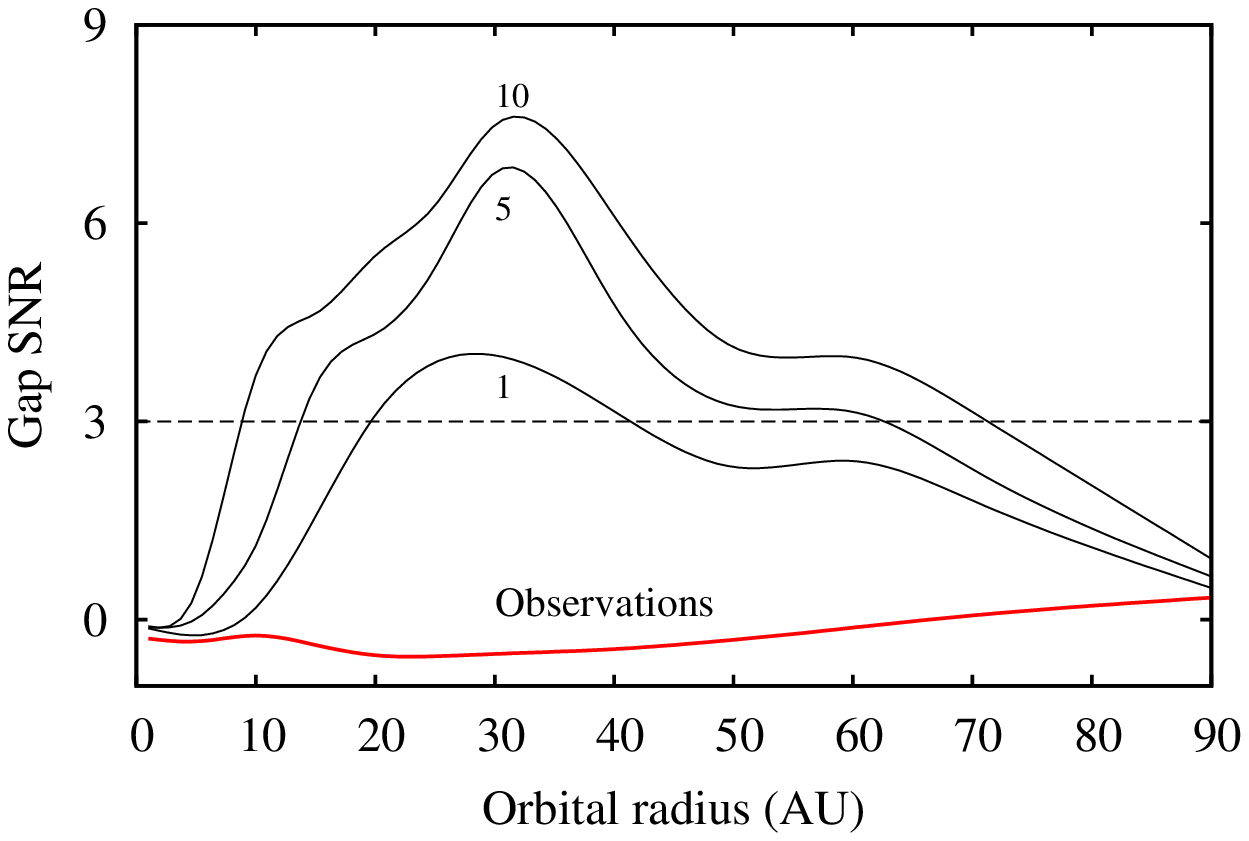}
\caption{\label{fig:planet_deviation} The signal-to-noise
ratio of the detection of a gap generated by a planet as a function of the orbital radius $R_p$ 
and the planet mass. The different curves correspond to masses between 0.5 and 10 Jupiter 
masses as labeled. The thick red curves indicate the signal-to-noise ratio measured from the 1.3 mm 
images after subtracting the best fit model for the similarity solution.} For {\it Gap SNR}$>3$, planets should 
produce a detectable gap. The upper and lower panel refers to the case of DG Tau and RY Tau respectively.

\end{figure}

The results are summarized in  the upper panel of Figure~\ref{fig:planet_deviation}. Planets 
with masses and radii that lead to {\it gap SNR}$>3$ 
produce detectable gaps. No gaps are detected in our observations of the DG~Tau disk at more 
than 3$\sigma$ (see the red curve). This enable us to constrain the masses and orbital radii of any 
planets that may be present. In particular, we can exclude that planets more massive than  
Jupiter exist between 5  and 40 AU, or that planets with masses slightly smaller than Jupiter exists between 
10 and 25 AU. The observations lack both the angular resolution and the 
sensitivity required to detect gaps produced by planets with a mass smaller 
than about 0.5~M$_J$.

\begin{figure}[t]
\centering
\includegraphics[angle=0,width=\columnwidth]{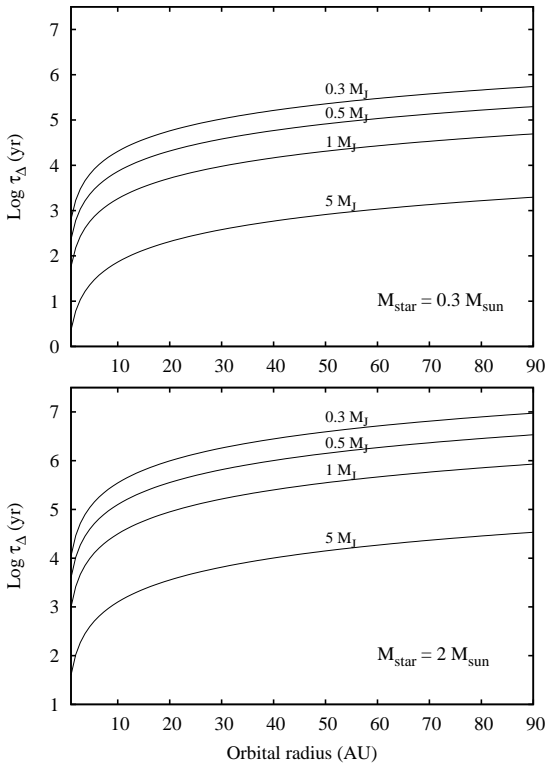}
\caption{\label{fig:tau_delta} Minimum time scale ($\tau_\Delta$)for the formation of a gap 
as a function of the orbital radius and mass of the planet. The different
curves correspond to planet masses between 0.3 and 5 M$_J$ as labeled 
in the figure. The upper and lower panels show the time scale
for a central star of 0.3~$M_{\sun}$ and 2~$M_{\sun}$ respectively.
}
\end{figure}

An important caveat is that a planet may exist but may not have had 
enough time to completely open a gap in the disk. The gap formation time 
scale $\tau_\Delta$ results from the trade-off between the efficiency of the 
tidal torque exercised by the planet in removing angular momentum, and the 
accretion of new material coming from larger radii in the gap due to the disk 
viscosity. A lower limit of the gap formation time scale is obtained in the zero 
viscosity limit. In this case, an analytic formulation is provided 
by \cite{br99} in the form 
\begin{equation}
\label{eq:taudelta}
\tau^{min}_\Delta  \simeq \frac{P}{q^2} \left( \frac{\Delta}{R_p} \right)^5,
\end{equation}
where $P$ is the orbital period, $q=M_p/M_{\star}$ and $\Delta = 2R_H$ as defined above. 
Assuming Keplerian rotation, we can 
rewrite the time scale for the gap formation as  
\begin{equation}
\tau^{min}_\Delta = 1.1 \, \textrm{Myr}\, \times \left(\frac{M_\star}{M_\sun}\right)^{3/2} 
\left(\frac{R_p}{\textrm{AU}}\right)^{3/2} \left(\frac{M_p}{M_J}\right)^{-2} 
\left(\frac{\Delta}{R_p}\right)^{5}
\end{equation}
The upper panel of Figure~\ref{fig:tau_delta} shows the calculated values 
of $\tau^{min}_\Delta$ for the stellar mass of DG~Tau (0.3~$M_{\sun}$). In the case 
of a planet with a mass between 0.3 and 0.5 M$_J$ orbiting at a radius larger 
than 40 AU, the minimum time scale for the gap formation is comparable with the age 
of the system (0.1~Myr). 
For more massive planets, or for 
closer radii, the minimum gaps time scale is a small fraction of the age of 
the system. 

We conclude that, for DG~Tau, the observations lack the sensitivity and 
angular resolution required to investigate the presence of planets 
less massive than about 0.5 M$_J$ at any orbital radius. Our analysis indicates that 
no planets more massive than Jupiter are present between 5 and 50 AU, 
unless they are younger than $10^4$ years.

\subsubsection{RY~Tau}
\label{sec:res_RYTau}
The similarity solution for the disk surface density is characterized 
by $\gamma= -0.56\pm0.18$ and $R_t\sim26\pm3$~AU. 
As shown in Figure~\ref{fig:sigma}, the surface density increases roughly 
as $\sqrt{R}$ from the inner radius at 0.1 AU up to about 26~AU 
and then decreases exponentially outward. This supports the suggestion
in Section~\ref{sec:morp} that the RY~Tau inner disk might be partially 
dust depleted with respect to power law disk models. 
We note that this surface density profile may provide an 
explanation for both the double peak intensity at 1.3 mm  and the disk excess at 
infrared wavelengths. Indeed, within 10 AU the model disk remains optically 
thick at optical and infrared wavelengths, exhibiting the infrared excess typical 
of classical disks. 

At larger radii, the surface density in RY Tau disk decreases 
smoothly and the residuals calculated by subtracting the best fit models 
to the 1.3~mm dust emission map do not show any structure at more 
than 3$\sigma$. This excludes strong deviations from an unperturbed viscous 
disk profile. 

The lower panel of Figure~\ref{fig:planet_deviation} shows the signal-to-noise ratio 
of the detection of a gap generated by planets of 1, 5 and 10 Jupiter masses as 
a function of the orbital radius. Due to the higher disk inclination and stellar mass, 
a planet orbiting around RY Tau would produce a less visible gap. In particular, 
our observations seems to exclude the presence of planets more massive than 5 
Jupiter masses between 10 and 60 AU. Given the higher stellar mass 
of RY~Tau, the minimum time scale for the formation of gaps is one order of 
magnitude larger than the case of DG~Tau (see lower panel of 
Figure~\ref{fig:tau_delta}). This implies that planets less massive than 
Jupiter orbiting at more than about 30 AU may not have had enough time to 
form a gap in the disk.

\subsection{Radial dependence of the dust properties}
\label{sec:betavar}

A comparison of the best fit solutions obtained for 
the wavelengths of 1.3~mm and 2.8~mm enable us to investigate the dependence 
of dust opacity on the orbital radius. If the dust opacity is constant throughout 
the disk as assumed in Section~\ref{sec:mod}, the model fitting necessarily 
leads to the same surface density profile for observations at two 
different wavelengths. Otherwise, different $\Sigma(R)$ would suggest a 
radial variation in the relative dust opacities at the observed wavelengths. 
To understand this point,  we assume that the dust emission
is optically thin. In this case the observations constrain the product 
$\Sigma_{\lambda}(R) \times k_{\lambda}$, where $\Sigma_{\lambda}(R)$ is the surface 
density obtained by fitting the observations at the wavelength $\lambda$.  
In the more general case in which the dust opacity depends on the orbital 
radius we can write
\begin{equation}
\label{eq:sigmak}
\Sigma_\lambda(R)\times k_\lambda = \tilde\Sigma(R)\times\tilde k_\lambda(R).
\end{equation} 
The left side of this equation contains the opacity discussed in 
Section~\ref{sec:mod} and the surface density derived from the model fitting.
The right side contains the unknown ``true'' surface 
density $\tilde\Sigma(R)$ in the case in which the ``true'' dust opacity 
$\tilde k_{\lambda}(R)$ varies with the radius. 
The ratio of Equation~\ref{eq:sigmak} for two different wavelengths
$\lambda_0$ and $\lambda_1$ leads to 
\begin{equation}
\frac{\Sigma_{\lambda_0}(R)}{\Sigma_{\lambda_1}(R)} \times \frac{k_{\lambda_0}}
{k_{\lambda_1}} = \frac{\tilde k_{\lambda_0}(R)}{\tilde k_{\lambda_1}(R)} = 
\left( \frac{\lambda_1}{\lambda_0} \right)^{\beta(R)} 
\end{equation}
Here we assumed that at each radius the dust opacity can be expressed by
a power law $k_{\lambda} \propto \lambda^{-\beta}$. Finally, taking the 
logarithm of this latter equation we can write
\begin{equation}
\label{eq:betar}
\beta(R) = \beta_c + \Delta\beta(R)
\end{equation}
where $\beta_c=\log(k_{\lambda_1}/k_{\lambda_0})/\log(\lambda_0/\lambda_1)$ 
and $\Delta\beta(R)$ has the form
\begin{equation}
\Delta\beta(R) = \log^{-1}\left( \frac{\lambda_1}{\lambda_0}\right) 
\times \log\left[ \frac{\Sigma_{\lambda_0}(R)}{\Sigma_{\lambda_1}(R)} \right].
\end{equation}
If $\Sigma_{\lambda_1}=\Sigma_{\lambda_0}$, the dust opacity slope is constant
throughout the disk and assumes the value discussed in Section~\ref{sec:mod}.
Otherwise, we can use the latter equation to investigate the radial variation 
of $\beta$.

The best fit solutions for $\Sigma(R)$ obtained at 1.3~mm and 2.8~mm are shown
in Figure~\ref{fig:RYTau_POW}-\ref{fig:DGTau_SIM} with black and red curves 
respectively. The best fit parameters 
are summarized in Table~\ref{tab:res_clubs} and \ref{tab:res_spades}.
The quoted uncertainties correspond to a likelihood of 99.7\% (i.e. 3$\sigma$)
 and are calculated by fitting a normal distribution to the measured 
probability distributions. For RY~Tau, the disk model obtained by fitting the 
two wavelengths separately are in agreement within 3$\sigma$. 
For DG~Tau the solutions disagree by more than 3$\sigma$ only in the case of 
the similarity solution and high dust opacity. 

\begin{figure*}
\centering
\includegraphics[angle=0,width=\textwidth]{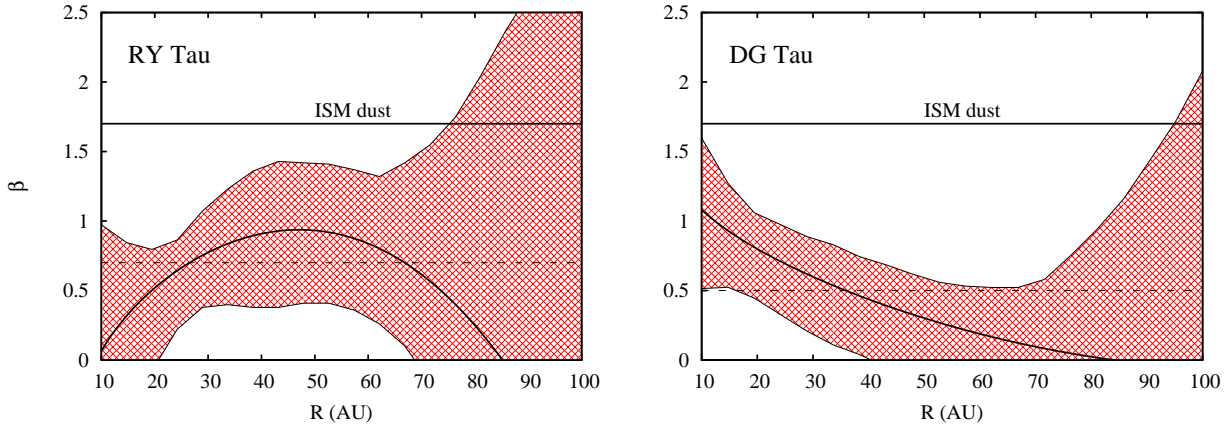}
\caption{\label{fig:dbeta} Slope of the millimeter dust opacity $\beta$ as 
a function of the radius for RY~Tau (left panel) and DG~Tau (right panel).
Values of $\beta$ outside the colored region are excluded by our observations at 
more than the 3$\sigma$ level. The dashed lines show the mean value of $\beta$ derived
in Paper I as discussed in Section~\ref{sec:mod}. The thick solid 
curve shows the radial profile of $\beta$ corresponding to the best fit
solution for the disk surface density at both 1.3 and 2.8~mm. The solid horizontal line 
shows the value of $\beta$ characteristic of the dust in the interstellar medium.}
\end{figure*}

Figure~\ref{fig:dbeta} shows the radial variation of $\beta$ as 
defined in Equation~\ref{eq:betar} for both DG Tau and RY Tau. 
The region marked with color indicates values 
of $\beta$ within 3$\sigma$ from the radial profile corresponding 
to the best fit solution for the surface density in the case of the 
similarity solution model. Values of $\beta$ outside 
this region are rejected by our observations. 

The results for RY~Tau and DG~Tau are clearly consistent with a large variety 
of radial profiles of $\beta$.  For both sources, $\beta$ is 
better constrained between radii of 20 and 70~AU. However, 
even in this interval, the observations constrain possible variation of $\beta$ to 
within only  $\Delta\beta < 0.7$. Nevertheless,  
across most of the disk the circumstellar dust differs from 
that observed in the interstellar medium (ISM). Dust in the ISM 
is characterized by sub-micron dust grains and by a millimeter opacity 
slope of $\beta \sim 1.7$. In contrast, in both sources,
$\beta$ is smaller than 1.7 up to at least 80 AU, suggesting that the 
circumstellar dust has been processed and, in particular, has 
increased its size up to a maximum value that varies between 20 $\mu$m
and a few centimeters.

Although in both sources $\beta$ may be constant throughout the disk (see
dashed lines), our results suggest that $\beta$ 
decreases with the radius in DG Tau disk. As discussed in Section~\ref{sec:mod}, 
$\beta$ depends on a number of poorly constrained quantities, such as 
composition, structure, and size of the dust grains. For example, 
varying $\beta$ from $\sim$1 to $\sim$0.2, similar to 
what is suggested for DG~Tau between 10 and 60 AU, may be due to the maximum 
grain size increasing from $20$~$\mu$m to 1~cm for $q=3$, or, 
alternatively, to a decrease of $q$ from 4 to 3 if the maximum grain size
is between 1 and 10~cm. In short, the interpretation 
of $\beta$ only in terms of the grain size distribution can be very 
misleading. It seems most plausible that both
dust composition and the relative contributions of smaller and larger 
grains change through the disk, contributing to the variation 
of the dust opacity. It is clear that better constraints on 
the radial profile of $\beta$ are required before pushing the 
investigation further.

In this regard, we note that the current constraints on the radial variation of
$\beta$ are limited by two factors. First, although the angular resolution of 
the observations described here is significantly better than hitherto possible, 
the dust surface density is well constrained only between 15 and 50 AU, 
where most of the observed flux is emitted (see the discussion in Section~\ref{sec:surf}).
At smaller and larger radii the surface density is uncertain by almost
one order of magnitude. Second, our analysis is hampered by the small 
separation in wavelength between the observations since $\Delta\beta(R)$ 
is proportional to $log^{-1}(\lambda_1/\lambda_0)$. The uncertainties 
shown in Figure~\ref{fig:dbeta} can be reduced by a factor of 2 by extending 
the observations at 7~mm. These observations will become possible with the 
expanded correlator on the EVLA.

\section{Conclusions}
\label{sec:conc}

We have presented CARMA observations of the dust thermal emission at
the wavelengths of 1.3~mm and 2.8~mm from the  circumstellar
disks around the pre-main sequence stars RY~Tau and DG~Tau. 
The observations are characterized by  unprecedented angular 
resolution of $\sim$0.15\arcsec\ and 0.30\arcsec\ at 1.3~mm 
and 2.8~mm respectively, corresponding to spatial scales of 20 
and 40 AU at the distance of Taurus. Based on these images, 
we have addressed three fundamental questions related to the formation 
of planets in the disk around pre-main sequence stars. What is the radial 
density distribution of circumstellar dust? Does the dust emission 
show any indication of the presence of (proto)-planets? Do the dust properties 
vary with orbital radius?

By analyzing the morphology of the surface brightness of the dust emission 
and comparing the observations with theoretical disk models, we make the 
following conclusions:

\begin{itemize}

\item[(1)]  Both the classical power law disk surface density (Hayashi 1981) 
and the similarity solution for the viscous evolution of a Keplerian 
disk (Hartmann 1998) fit the observations well. The surface 
density is well constrained between 15 and 50 AU. In this region, 
the two models lead to values of $\Sigma$ that agree within 30\% for a fixed dust opacity. 
At smaller and larger radii, the surface density depends on the 
assumed model and varies by almost an order of magnitude. We have 
verified that the assumptions on the dust opacity have a small effect on the 
model fitting and, therefore, on the radial
profile of the dust density. However, the total disk mass may vary 
by almost two order of magnitude for different dust compositions and 
grain size distributions.

\item[(2)] The dust emission in DG~Tau is mostly radially symmetric. 
It is characterized by a single, central peak and smoothly decreases up to
an angular distance of about 0.5\arcsec. Theoretical disk models  
reproduce the observation very well, with randomly distributed 
residuals between 3 and 6$\sigma$. No systematic deviation from the similarity 
solution for the surface density of a viscous disk are observed. By simulating 
the presence of planets in the disk via the gap in the surface density 
produced by tidal torques, we find that the observations exclude the 
presence of planets more massive than Jupiter orbiting between 5 and 40 AU
from the central star, unless the planets are very young ($<10^4$ yr) and 
have not had the time to open a gap in the disk. The observations 
lack both the angular resolution and sensitivity to investigate the 
presence of planets less massive than about 0.5~Jupiter masses. 

For RY~Tau, the dust emission is characterized by two peaks separated 
by about 28~AU that suggest a decrease in the surface density, or dust opacity, 
within 14 AU of the central star. We found that the similarity solution for the disk surface 
density is characterized by a negative value of $\gamma$, and provides a reasonable explanation 
for the double peak intensity observed at 1.3 mm. Depletion of millimeter 
dust grains \citep{dd05}, decreasing values of the disk viscosity, or the
presence of planetesimals, may produce the observed
dust morphology. At larger radii, the dust emission shows a very smooth profile with 
no asymmetries or gaps. The lack of gaps in the disk suggests that any planets between 
10 and 50 AU are less massive than about 5 Jupiter masses, or, as for DG Tau, are very young.

\item[(3)] The best-fit models to the 1.3~mm and 2.8~mm data were compared 
to investigate the radial dependence 
of the slope opacity $\beta$, assuming that the dust opacity at millimeter 
wavelengths is expressed by a power law $k_\lambda \propto \lambda^{-\beta}$. 
We can exclude cases in which $\beta$ varies by more than 0.7 within 70 AU. 
Nevertheless, between 20 and 70 AU, the disks around DG~Tau and RY~Tau 
are characterized by values of $\beta$ that are smaller than that found in the ISM.
This implies that the dust has been reprocessed and has grown in size up
to a radius of at least 20 $\mu$m. The investigation of the radial 
variation of $\beta$ is still limited by 
the angular resolution and by the small separation in wavelength between 
the observations. In the future, ALMA and the EVLA will play 
crucial roles in the investigation of the radial 
dependence of the dust properties by increasing  the angular resolution 
and the interval in wavelength.
\end{itemize}

\acknowledgments

We thank Antonella Natta and Steven Beckwith for reading the manuscript and providing 
helpful suggestions. We thank the OVRO/CARMA staff and the CARMA observers 
for their assistance in obtaining the data. Support for the 
Combined Array for Research in Millimeter Astronomy
construction was derived from the Gordon and Betty Moore Foundation, 
the Kenneth T. and Eileen L. Norris Foundation, the Associates of the 
California Institute of Technology, the states of California, Illinois, 
 and Maryland, and the National Science Foundation. Ongoing CARMA 
development and operations are supported by the National Science 
Foundation  under a cooperative agreement (grant AST 08-38260), and by the 
CARMA partner universities. This work was performed in part under 
contract with the Jet Propulsion Laboratory (JPL) funded by NASA through the 
Michelson Fellowship Program. JPL is managed for NASA by the California 
Institute of Technology.

\appendix
\section{Effects of the assumptions on the dust composition on the dust opacity and size}
\label{app:A}

The slope of the dust opacity $\beta$ measured at millimeter wavelengths
has been widely adopted to constrain the size and opacity of the circumstellar  
dust \citep[see, e.g.,][]{r09}.  In this appendix we investigate how these latter two quantities 
depend on the assumption on the grain size distribution and composition.  

\begin{figure}[!t]
\centering
\includegraphics[angle=0,width=7.5cm]{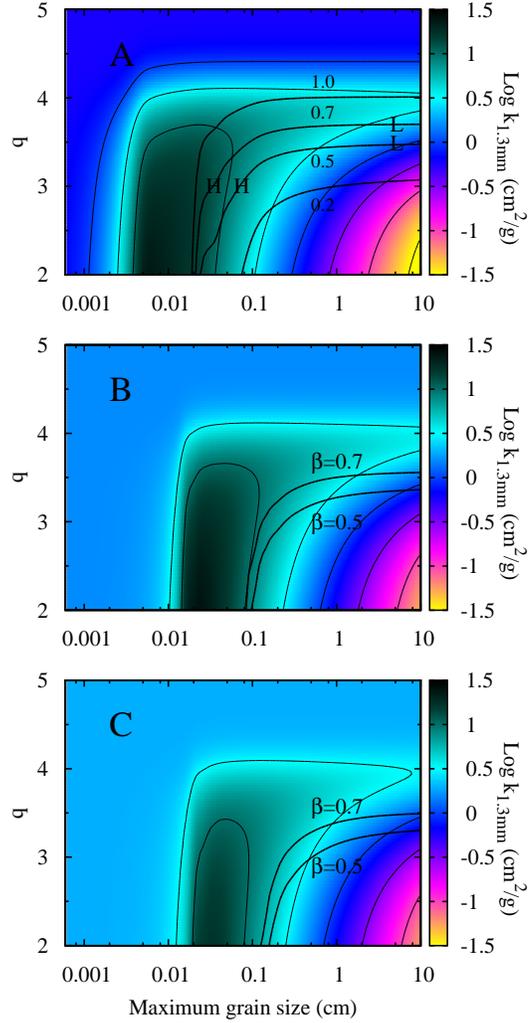}
\caption{\label{fig:k_beta} The color gradient depicts the dust opacity 
as a function of the maximum grain size $a_{max}$ and the slope of the 
grain size distribution $q$. Results are shown for three different dust models (A, B and C) as described in the text. 
The minimum grain size is fixed at 
$5\times10^{-7}$~cm. Thin solid curves show the dust opacity 
contours and are spaced by 0.5 dex. The thick solid curves show 
the possible pairs of $(a_{max},q)$ that lead to values 
for the opacity slope $\beta$ equal to 0.5 (DG~Tau) 
and 0.7~(RY~Tau). The letters $H$ and $L$ indicate the high 
and low dust opacity models adopted in fitting the observed dust 
emission. 
}
\end{figure}

The color scale in Figure~\ref{fig:k_beta} shows the dust opacity at 1.3 mm  
calculated for three different dust compositions.  In model A we assume that the 
grains are compact spheres composed of astronomical silicates and organic
carbonates \citep{wd01,zu96}.  This is the dust model adopted in the paper. 
In model B the grain is composed of silicates, carbonates and water ice. Finally, 
in model C the ice is replaced by a vacuum,  resulting in a porous grain made of  
silicates and carbonates. We assume a mass ratio of 1 between silicates and 
organics, and of 0.7  between silicates and the water ice or the vacuum \citep{po04}. 
The resulting bulk densities are 2.5 g/cm$^3$ for model A, 1.9 g/cm$^3$ for 
model B and 1.5 g/cm$^3$ for model C. We assume a grain size distribution 
$n(a) \propto a^{-q}$ and fix the minimum grain size to 0.005 $\mu$m. We then 
calculate the dust opacity by varying the maximum grain size $a_{max}$  
between $6\times10^{-4}$ and 10~cm and the slope $q$ between 2 and 5.
  
The solid thick curves show the values of $(a_{max},q)$ required to obtain the 
values of $\beta$  measured for DG Tau and RY Tau, 0.5 and 0.7 respectively. 
From the figure, it is clear that the value of $\beta$ sets a lower bound to the 
maximum grain radius in the grain size distribution \citep[see, e.g.,][]{nat04}. 
This lower bound is a function of the grain composition and increases by almost 
an order of magnitude between model A and C. For example, in order to 
have $\beta = 0.7$, the maximum grain size must be at least $0.03$ cm for 
compact grains in model A, or about 0.2~cm for the porous grains in model C. 
Note, however, that we can obtain the same value of $\beta$ with $a_{max}=10$ 
cm and $q$ between 3.2 and 3.7.  Deriving the maximum grain size from the 
measure of $\beta$ is evidently strongly degenerate. 

This introduces a large uncertainty on the dust opacity and ultimately
on the total mass of circumstellar dust derived from millimeter observations. 
Even if we limit the analysis to the generally adopted value $q=3.5$ \citep[see, e.g.,][]{bdh08}, 
the dust opacity for $\beta=0.7$ varies from about 8.4 cm$^2$/g of dust in 
the case of the model A and $a_{max}=0.1$ cm, to 1.2~cm$^2$/g of dust for 
model C and $a_{max}=10$~cm, leading to disk masses that differ by almost one order
of magnitude.

\end{document}